\documentclass[]{aa}  
\usepackage{amsmath}
\usepackage{amssymb}
\usepackage{color}
\usepackage{url}
\usepackage{ulem}
\usepackage{multirow}
\usepackage{graphics}
\usepackage{longtable}
\usepackage{graphicx}
\usepackage{bm}
\usepackage{appendix}
\usepackage{lastpage}

\usepackage{txfonts}

\DeclareGraphicsExtensions{.pdf,.jpeg,.png,.jpg,.eps}

\usepackage{wasysym}

\usepackage{twoopt}
\usepackage[breaklinks=true,pagebackref=false,colorlinks=true,citecolor=blue]{hyperref} 
\bibpunct{(}{)}{;}{a}{}{,} 

\makeatletter
\renewcommand*\aa@pageof{, page~\thepage{} of~\pageref*{LastPage}}
 
\makeatother

\begin{document}

\authorrunning{D.-C. Chen et al.}
\title{The New Generation Planetary Population Synthesis (NGPPS)} \subtitle{\uppercase\expandafter{\romannumeral8}. Impact of host star metallicity on planet occurrence rates, orbital periods, eccentricities,  and radius valley morphology}
\titlerunning{The New Generation Planetary Population Synthesis (NGPPS). VIII.}

\author{Di-Chang Chen
          \inst{1,2,3,4*}, Christoph Mordasini\inst{2,5}, Alexandre Emsenhuber\inst{2}, Remo Burn\inst{6},  Ji-Wei Xie\inst{3,4} \and Ji-Lin Zhou\inst{3,4}
          }
   \institute{School of Physics and Astronomy, Sun Yat-sen University, Zhuhai 519082, China, \email{chendch28@mail.sysu.edu.cn}
   \and
   Division of Space Research and Planetary Sciences, Physics Institute, University of Bern, Sidlerstrasse 5, 3012 Bern, Switzerland
   \and
   School of Astronomy and Space Science, Nanjing University, Nanjing 210023, China
   \and
   Key Laboratory of Modern Astronomy and Astrophysics, Ministry of Education, Nanjing 210023, China
   \and
   Center for Space and Habitability, University of Bern, Gesellschaftsstrasse 6, 3012 Bern, Switzerland
   \and
   Max-Planck-Institut f\"ur Astronomie, K\"onigstuhl 17, Heidelberg, 69117, Germany\\
   \renewcommand{\thefootnote}{}   
   }
   \date{Received ; accepted}

\abstract
{The dust-to-gas ratio in the protoplanetary disk, which is likely imprinted into the host star metallicity, is a property that plays a crucial role during planet formation. On the observational side, statistical studies based on large exoplanet datasets have determined various correlations between planetary characteristics and host star metallicity.} 
{We aim at constraining planet formation and evolution processes by statistically analysing planetary systems produced at different metallicities by a theoretical model, comparing with the correlations derived from observational samples.
}
{We use the Generation III Bern  model of planet formation and evolution to generate synthetic planetary systems at different metallicities.
This global model incorporates the accretion of planetesimals and gas, planetary migration, N-body interactions among embryos, giant impacts, protoplanetary disk evolution, as well as the planets' long-term contraction and atmospheric loss of gaseous envelopes. Using synthetic planets biased to observational completeness, we analyse the impact of stellar metallicity on planet occurrence rates, orbital periods, eccentricities, and the morphology of the radius valley.
}
{Based on our nominal model, we find that (1) the occurrence rates of large giant planets and Neptune-size planets are positively correlated with $\rm [Fe/H]$, while small sub-Earths exhibit an anti-correlation. In between, at radii of 1 to 3.5 $R_\oplus$, the occurrence rate first increases and then decreases with increasing $\rm [Fe/H]$ with an inflection point at $\sim$0.1 dex. (2) Planets with orbital periods shorter than ten days are more likely to be found around stars with higher metallicity, and this tendency weakens with increasing planet radius.
(3) Both giant planets and small planets exhibit a positive correlation between the eccentricity and $\rm [Fe/H]$, which could be explained by the self-excitation and perturbation of outer giant planets. (4) The radius valley deepens and becomes more prominent with increasing $\rm [Fe/H]$, accompanied by a lower super-Earth-to-sub-Neptune ratio. Furthermore, the average radius of the planets above the valley (2.1–6 $R_\oplus$) increases with $\rm [Fe/H]$.}
{Our nominal model successfully reproduces many observed correlations with stellar metallicity either quantitatively or qualitatively, supporting the description of physical processes and parameters included in the Bern model. 
Quantitatively, the dependences of orbital eccentricity and period on $\rm [Fe/H]$ predicted by the synthetic population is however significantly weaker than observed. This discrepancy likely arises because the  model accounts for planetary interactions only for the first 100 Myr and neglects the effects of the stellar environment (e.g. clusters, binaries). This suggests that long-term dynamical interactions between planets, along with the impact of binaries/companions, can drive the system towards a dynamically hotter state.
}

\maketitle

\section{Introduction} \label{sec:intro}
The number of discovered exoplanets {\color{black} has exceeded} 5500 through various surveys like HARPS \citep{2011arXiv1109.2497M}, Lick/Keck \citep{2014ApJS..210....5F,2017AJ....153..208B}, CoRoT \citep{2006MNRAS.372.1117C}, Kepler \citep{2010Sci...327..977B,2018ApJ...856...37B}, and TESS \citep{2015JATIS...1a4003R}.
Based on such a large observational dataset, numerous works have provided crucial constraints on the characteristics of exoplanetary systems \citep[e.g. radius, occurrence rate, eccentricity, inclination, and orbital period ratio;][]{2014ApJ...790..146F,2017AJ....154..109F,2018ARA&A..56..175D,2018ApJ...860..101Z}. 
Moreover, various correlations between the planetary characteristics and stellar properties \citep[see the review by][]{2021ARA&A..59..291Z} have been determined, which provide important {\color{black} insights into} the planet formation and evolution theory \citep{2018haex.bookE.153M}. 

Of various stellar properties, the stellar metallicity is a crucial factor that has important impacts on planet formation.
From theory, as a proxy to the dust-gas-ratio of the protoplanetary disk, together with the disk mass, the metallicity will determine the amount of solid materials in the disk and thus affect the planetary formation processes \citep{2021A&A...656A..69E}, including the growth from dust to planetesimals \citep{1993Icar..106..210I,2001Icar..152..205C,2010A&A...520A..43O}, the gas accretion \citep{1996Icar..124...62P,2000Icar..143....2B,2012MNRAS.427.2597A} and planet migration \citep{2014prpl.conf..667B,2016MNRAS.458.3927B}.
From observation, the stellar metallicity is found to influence various planetary characteristics (e.g. occurrence rate, orbital architecture).
For example, giant planets have been well established to be preferentially hosted by metal-richer stars \citep{2001A&A...373.1019S,2005ApJ...622.1102F,2015AJ....149...14W,2023PNAS..12004179C}.
Some recent studies also show that the occurrence rate of smaller planets (especially in short-period orbits) may also be correlated with stellar metallicity \citep{2015AJ....149...14W,2016ApJ...832...34M,2018PNAS..115..266D,2019ApJ...873....8Z}.
The orbital eccentricity is also found to be positively-correlated with the stellar metallicity both for giant planets \citep{2013ApJ...775..105O,2018ApJ...856...37B} and small planets as found by the Kepler mission \citep{2019AJ....157..198M,2023AJ....165..125A}. 

A powerful approach to constrain the key process of planetary formation and evolution is to perform planetary population synthesis \citep{2004ApJ...616..567I,2009A&A...501.1161M,2024arXiv241000093B} for statistical comparisons between the produced planetary systems of different metallicity from the theoretical model and the observed exoplanetary systems.
This requires a global formation and evolution model \citep{2015IJAsB..14..201M,2022ASSL..466....3R} that can predict the properties of planetary systems from different initial conditions (e.g. disk mass, dust-gas-ratio) that are representative of known protoplanetary disks.
Furthermore, the detection bias of observational data should also be applied to the synthetic populations.

Here we rely on the Generation \uppercase\expandafter{\romannumeral3} Bern model, a combined global end-end planetary formation and evolution model \citep{2021A&A...656A..69E}, that was presented in the NGPPS I paper.
By adopting the Bern model, we can generate synthetic planet populations which can predict various properties (e.g. mass, radii, and orbits) for planets of different sizes.
In \citet[][NGPPS II]{2021A&A...656A..70E}, we used this model to perform the New Generation Planetary Population Synthesis of multi-planetary systems and discussed the initial conditions so that they can be representative of the protoplanetary disk population. 
For the detection biases, in \citet[][NGPPS VI]{2021A&A...656A..74M}, we introduced the KOBE program to simulate the geometrical limitations of the transit method and apply the detection biases and completeness of the Kepler survey.
In \citet[][NGPPS VII]{NGPPSVII}, we also show how to apply the detection biases of radial velocity method to the synthetic population.
We have already performed analyses on the super Earth-cold Jupiters correlation \citep[][NGPPS III]{2021A&A...656A..71S} and the architecture of multiple transiting planetary systems \citep[][NGPPS VI]{2021A&A...656A..74M} using everywhere the same synthetic population.
Quantitative comparisons with various surveys (e.g. Kepler and HARPS) have also been performed \citep{2019ApJ...887..157M,2024A&A...687A..25C,NGPPSVII}.

In this paper, we continue the NGPPS series and make statistical analyses as a function of stellar metallicities using the synthetic population of solar-like stars and compare with the observational results.
The paper is organised as follows:
In Sect. \ref{sec:model}, we briefly introduce the coupled formation and evolution model.
In Sect. \ref{sec:Synthesis}, we describe how we set the initial conditions for our population synthesis.
Then we investigate the correlations between the stellar metallicity and the planetary occurrence rates (Sect. \ref{sec.frequency}), the radius valley morphology (Sect. \ref{sec.valley}), the orbital period (Sect. \ref{sec.distribution_radii}) and eccentricity (Sect. \ref{sec.eccen}) and make quantitatively comparisons with observations.
Finally, we summarise the main results in Sect. \ref{sec.Summary}.

\section{Formation and evolution model} \label{sec:model}

In this work, we use the Generation \uppercase\expandafter{\romannumeral3} Bern model of planetary formation and evolution, the latest generation of models first presented in \citet{2005A&A...434..343A}. A full description of the Generation \uppercase\expandafter{\romannumeral3} model was given in \citet{2021A&A...656A..69E}; therefore we will provide here only a brief overview. The Bern model is a global model for the formation and evolution of planetary systems combining self-consistently descriptions for a significant number of physical processes.

The model begins with the formation stage. There, the evolution of the protoplanetary disk, the dynamical state of the planetesimal disk, concurrent solid and gas accretion by the protoplanets, gas-driven migration, and dynamical interactions between the protoplanets are tracked together.

The disks have a 1D radial, axis-symmetric representation. The evolution of the gas disk is computed by solving the advection-diffusion equation \citep{1952ZNatA...7...87L,1974MNRAS.168..603L} with the turbulent viscosity set following the standard $\alpha$ parametrization \citep{1973A&A....24..337S}. The gas disk's vertical structure is computed following \citet{1994ApJ...421..640N}. Additional sink terms are included for the internal \citep{2001MNRAS.328..485C} and external \citep{2003ApJ...582..893M} photo-evaporation, plus accretion by the protoplanets. The planetesimal disk represents the solid component, with its dynamical state evolved following the balance between damping from the gas disk and stirring by the other planetesimals and the protoplanets \citep{2004AJ....128.1348R,2006Icar..180..496C,2013A&A...549A..44F}. 

Planet formation is assumed to occur by core accretion \citep{1974Icar...22..416P,1980PThPh..64..544M}. At the beginning, a given number of protoplanets with a mass of $10^{-2}\,M_\oplus$ are inserted in the disk. The core grows by planetesimal accretion, which is taken to be in the oligarchic regime \citep{1993Icar..106..210I,2003Icar..161..431T,2013A&A...549A..44F}. 
{\color{black} The planets are assumed to have an onion-like spherically symmetric structure with an iron core, a silicate mantle, a water ice layer and a gaseous envelope made of pure H/He. 
The internal structure of the planets (and thus their gas accretion rate, radius, luminosity, and interior structure) is calculated at all stages (attached, detached, evolution) by directly solving the 1D structure equations \citep{1986Icar...67..391B}.
For the composition of the planet core, we now use self-consistently the iron mass fraction as given by the disc compositional model according to \citet{2014A&A...562A..27T,2014A&A...570A..35M}.
This composition is tracked into the protoplanets when a propotoplanet accretes planetesimals, and in giant impacts between protoplanets, yielding to a final iron to silicate ratio.}
For the gas envelope structure, at early times, gas accretion is limited by planet's ability to radiate away the energy released by the accretion of both solid and gas. The efficiency of this process increases as the planet grows while the supply from the disk decreases as the disk depletes. Hence the gas accretion becomes limited by the supply from the disk at some point. When this occurs, the structure equations are used to determine the planet's radius, allowing to track the contraction and cooling \citep{2000Icar..143....2B}.

The model also includes the \texttt{Mercury} \textit{N}-body package \citep{1999MNRAS.304..793C} to track the dynamical interactions between the protoplanets. Gas-driven migration is included by means of additional forces in the \textit{N}-body. We follow the prescription of \citet{2014MNRAS.445..479C} for Type~I migration, and that of \citet{2014A&A...567A.121D} for Type~II migration. The transition between the two regimes is given by the criterion of \citet{2006Icar..181..587C}.

After a predetermined time of 100~Myr, the model switches to the evolution stage. Here, the model evolves each planet individually to 10~Gyr. The main aim is to track the thermodynamical evolution (cooling and contraction) of the planets following \citet{2012A&A...547A.111M}. This stage also includes atmospheric escape \citep{2014ApJ...795...65J,2018ApJ...853..163J,2024NatAs...8..463B}, and tidal migration \citep{2008CeMDA.101..171F,2009ApJ...698.1357J,2011A&A...528A...2B}. An important assumption, for planets whose mass is dominated by accreted solids, is how accreted volatile ices (mainly H$_2$O, CO$_2$, CH$_4$, CH$_3$OH, NH$_3$, but the model also allows for accretion of icy CO and N$_2$ in the very cold regions) are treated. Here, it is assumed that the mass which was accreted as ice evaporates and mixes perfectly with any hydrogen and helium which was accreted from the gas phase. Under this assumption, the mean molecular weight of the atmosphere increases, typically reducing the size of a H/He-rich planet but increasing the size for planets which would not contain H/He. To solve the internal structure of the mixed atmosphere, an equation of state (EOS) and opacities in the gaseous envelope are required. For both ingredients, a fully consistent treatment is challenging due to availability (EOS) or computational cost (opacity). Therefore, all volatile species are treated as H$_2$O for the EOS using AQUA \citep{2020A&A...643A.105H} which is combined using the additive volume law with the EOS for H/He \citep{2019ApJ...872...51C}. Given the high temperatures of observable planets, water is typically in the gas phase and expected to be mixed well with H/He which holds even for the escaping material in the surface layer \citep[see discussions in][]{2024NatAs...8..463B,2024arXiv241116879B}. For the opacity in the envelope, we assume a mixture of elements combining in chemical equilibrium to molecules assuming Solar abundances scaled with the overall metallicity \citep{2014ApJS..214...25F}.

For photo-evaporative mass loss, the approach of \citet{2024NatAs...8..463B} uses a mass-weighted average of the expected mass loss of H/He and pure H$_2$O following the works of \citet{2021RNAAS...5...74K,2018A&A...619A.151K} and \citep{2020ApJ...890...79J} respectively. For H$_2$O, this approach relies on Energy-limited escape using a fitted efficiency while the H/He escape is directly taken from tables based on hydrodynamic models.

\section{Population synthesis} 
\label{sec:Synthesis}

\begin{figure}[!t]
\centering
\includegraphics[width=\linewidth]{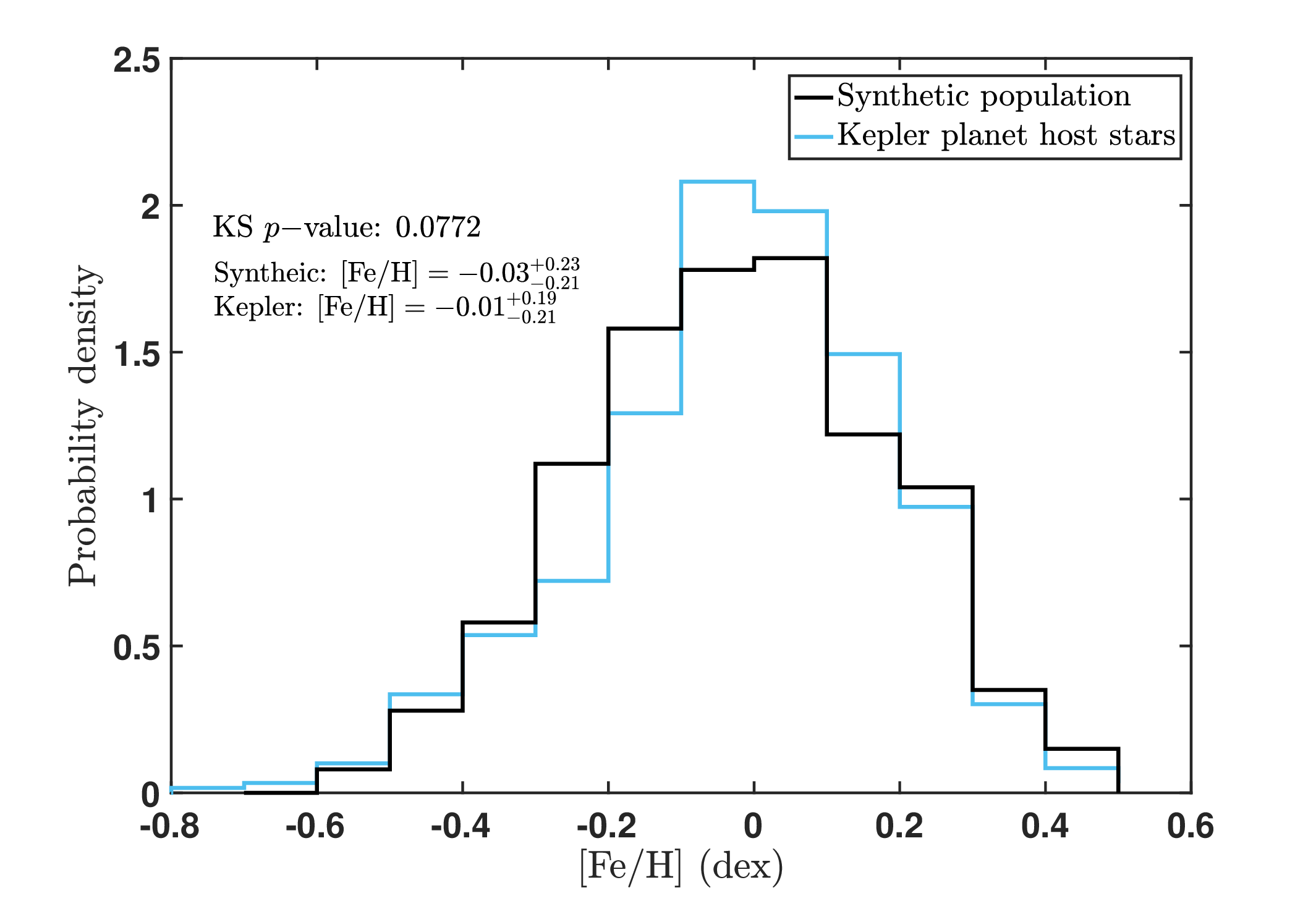}
\caption{Probability density functions of the stellar metallicity ($\rm [Fe/H]$) for the synthetic population (black) and Kepler Sun-like planet-host stars (cyan).
In the top-left corner, we also print the KS $p-$value and the median values (1-$\sigma$ interval) of the two samples.
\label{figStarFeHBernKepler}}
\end{figure}

A synthetic population of planetary systems is obtained by varying the initial conditions of the model. The procedure was described in \citet{2021A&A...656A..70E}. Here, we analyse data from the population with the internal identifier \texttt{NG76Longshot}. It is based on \texttt{NG76} as presented in \citet{2021A&A...656A..70E} but with prolonged N-body integration and modified assumptions for the long-term evolution \citep{2024NatAs...8..463B} as described in Sect. \ref{sec:model}. The simulation was first analysed with a focus on the radius distribution and labelled the \texttt{Mixed} case in \citet{2024NatAs...8..463B}.

The starting point of the simulation mimics a disk in the state after the collapse of a cloud in the Class I stage. At this time zero, we assume here that pebbles have drifted and planetesimals and 100 larger, 0.01\,M$_\oplus$-mass embryos have formed. The latter are distributed randomly in the disk following a uniform distribution in logarithmic distance to the star. To generate a population of synthetic planets, we simulated 1000 systems with varied disk properties of young protoplanetary disks derived from observation: their initial gas mass \citep{2018ApJS..238...19T}, metallicity, size \citep{2010ApJ...723.1241A}, inner edge \citep{2017A&A...599A..23V}, and lifetime (distributed around 3\,Myr by means of tuning the external photoevaporation rate). 
{\color{black} The initial solid mass of the  disc is calculated by multiplying the mass of gas $M_{\rm g}$} and dust-to-gas-ratio $f_{D/G}$.
For the $f_{D/G}$, we assume that stellar and disc metallicities are identical and thus,
\begin{equation}
    \frac{f_{D/G}}{f_{D/G,\oplus}} = 10^{\rm [Fe/H]}.
\end{equation}
The dust-to-gas of the Sun $f_{D/G,\oplus}$ is taken as 0.0149 \citep{2003ApJ...591.1220L}.
{\color{black} Figure \ref{figStarFeHBernKepler} displays the distribution of $\rm [Fe/H]$ for our synthetic population. 
As can be seen, the range of $\rm [Fe/H]$ is restricted to -0.6 to 0.5, with a median value similar to the solar metallicity (black line).
To compare with the observational sample, we select planet host stars as single Sun-like stars from the LAMOST-Kepler-Gaia catalogue \citep{2021AJ....162..100C} with the following criteria: $4700 {\rm K} < T_{\rm eff} < 6500 {\rm K}$ \& $\log g>4.0$ \& $\rm RUWE<1.2$.
The median value and 1-$\sigma$ range of $[\rm Fe/H]$ of the selected Kepler sample ($-0.01^{+0.19}_{-0.21}$) are similar to those of the synthetic population ($-0.03^{+0.23}_{-0.21}$).
We also perform two-sample Kolmogorov-Smirnov (KS) and the resulted $p-$value is 0.0772, suggesting that the metallicities of the two samples show no significant difference.}
Other model parameters are kept constant across the systems, such as the turbulent viscosity parameter $\alpha=2\times10^{-3}$ or the planetesimal size $R_{\rm plts} = 300\,$m.




\section{Occurrence rate vs. metallicity}
\label{sec.frequency}
The occurrence rate, i.e. the average number of planets per star, is one of the fundamental characteristics of exoplanet science.
Determining how the occurrence rate depends on the properties of the host stars can provide crucial insights into the planet formation and evolution \citep{2007ARA&A..45..397U,2021ARA&A..59..291Z}.
In this section, we divide the synthetic population into several bins according to their metallicities $\rm [Fe/H]$ and make quantitative analyses on the correlations between the occurrence rate of planets of different radii and the stellar $\rm [Fe/H]$.

\subsection{Giant planets}
\label{sec.frequency.GP}

\begin{figure*}[!t]
\centering
\includegraphics[width=0.9\linewidth]{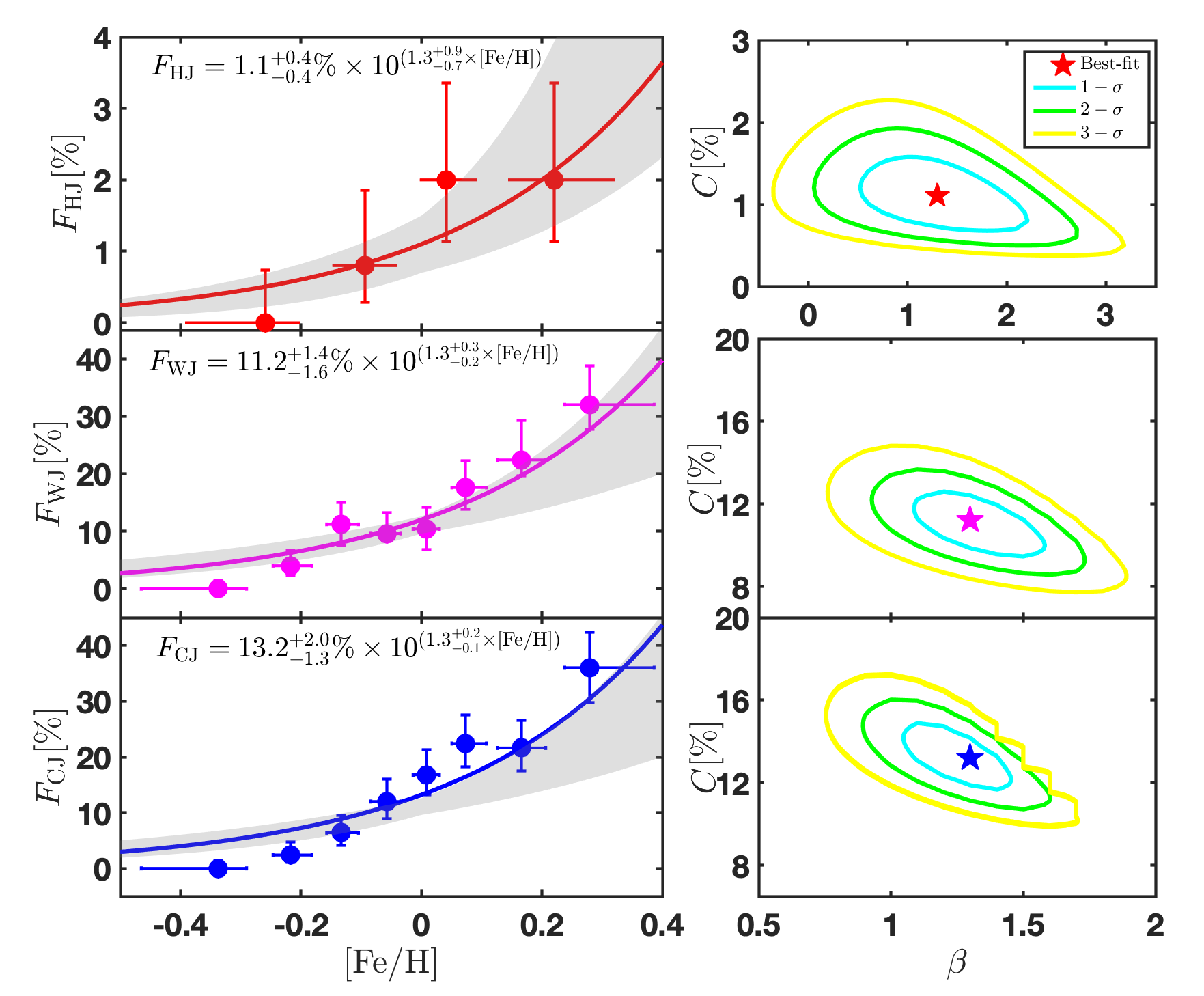}
\caption{Left panels: The occurrence rate of hot Jupiters (Top), warm Jupiters (Middle) and cold Jupiters (Bottom) as a function of stellar metallicity $\rm [Fe/H]$ from the synthetic NGPPS population and evolution model at 5 Gyr.
The solid lines denote the best-fits of Equation (1).
For the occurrence rate, we are more concerned with the increasing trend rather than the absolute magnitude. 
Thus, to compare the synthetic results with observations, we use the fitted $\beta$ derived from the observational sample in \citet{2023PNAS..12004179C} and the $C$ derived from the synthetic sample to generate the $1-\sigma$ intervals for the observational results, which are plotted as grey regions.
Right panels: Marginalized posterior probability density distributions of the parameters $(C, \beta)$ for the occurrence rates of hot Jupiters (Top), warm Jupiters (Middle) and cold Jupiters (Bottom) conditioned on the synthetic data generated by the Bern model.
\label{figFHJCJWJ_FeH_Synthesis}}
\end{figure*}

We initialize our giant planetary sample from the synthetic planet population and select planets with
masses between 0.1-13 Jupiter masses ($M_{\rm J}$) as giant planets.
According to previous studies \citep[e.g.][]{2018ARA&A..56..175D}, we further divide the giant planetary sample into three categories according to their orbital periods:
\begin{enumerate}
\item $P<10$ days, 12 hot Jupiters (HJs)

\item $10 \le \rm days \le P < 300$ days, 130 warm Jupiter (WJs)

\item $300 \ \rm days \le P \le 3300 \ \rm days$, 154 cold Jupiter (CJs).
\end{enumerate}
Here we set the typical time baselines (3300 days) for the observed time baseline for radius velocity (RV) surveys \citep[e.g. the LCES HIRES/Keck Precision Radial Velocity Exoplanet Survey, the public HARPS RV database;][]{2008psa..conf..287F,2014AN....335...41T,2017AJ....153..208B,2020A&A...636A..74T} as the upper limit of the period of cold Jupiters to facilitate subsequent comparison with observations.

\begin{figure*}[!t]
\centering
\includegraphics[width=0.9\linewidth]{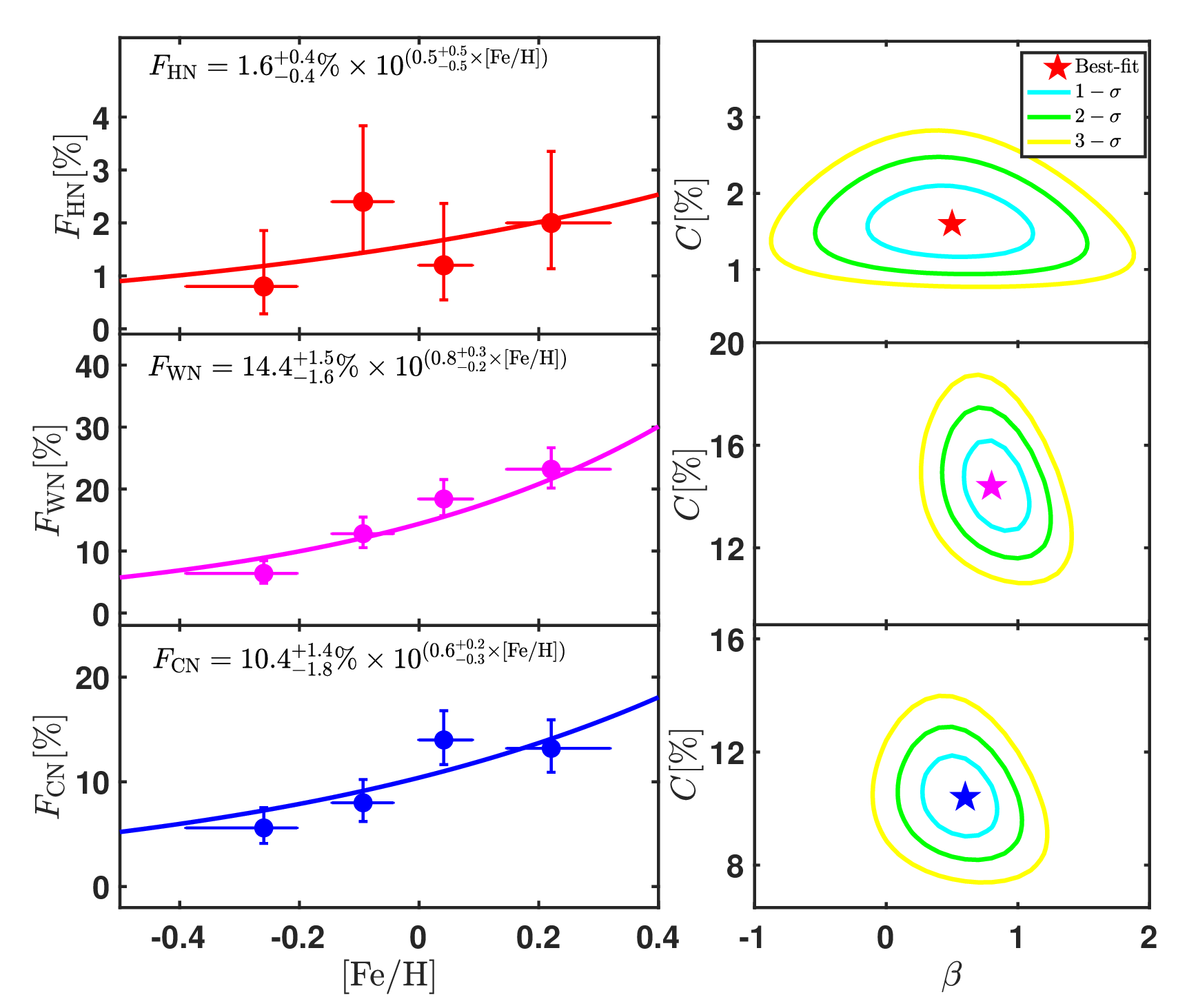}
\caption{Left panels: The occurrence rate of hot Neptunes (Top), warm Neptunes (Middle) and cold Neptunes (Bottom) as a function of stellar metallicity $\rm [Fe/H]$ from the synthetic NGPPS population by Bern planet formation and evolution model at 5 Gyr.
The solid lines denote the best-fits of Equation (1).
Right panels: Marginalized posterior probability density distributions of the parameters $(C, \beta)$ for the occurrence rates of hot Neptunes (Top), warm Neptunes (Middle) and cold Neptunes (Bottom) conditioned on the synthetic data generated by the Generation III Bern model.
\label{figFHNCNWN_FeH_Synthesis}}
\end{figure*}

\begin{figure*}[!t]
\centering
\includegraphics[width=\textwidth]{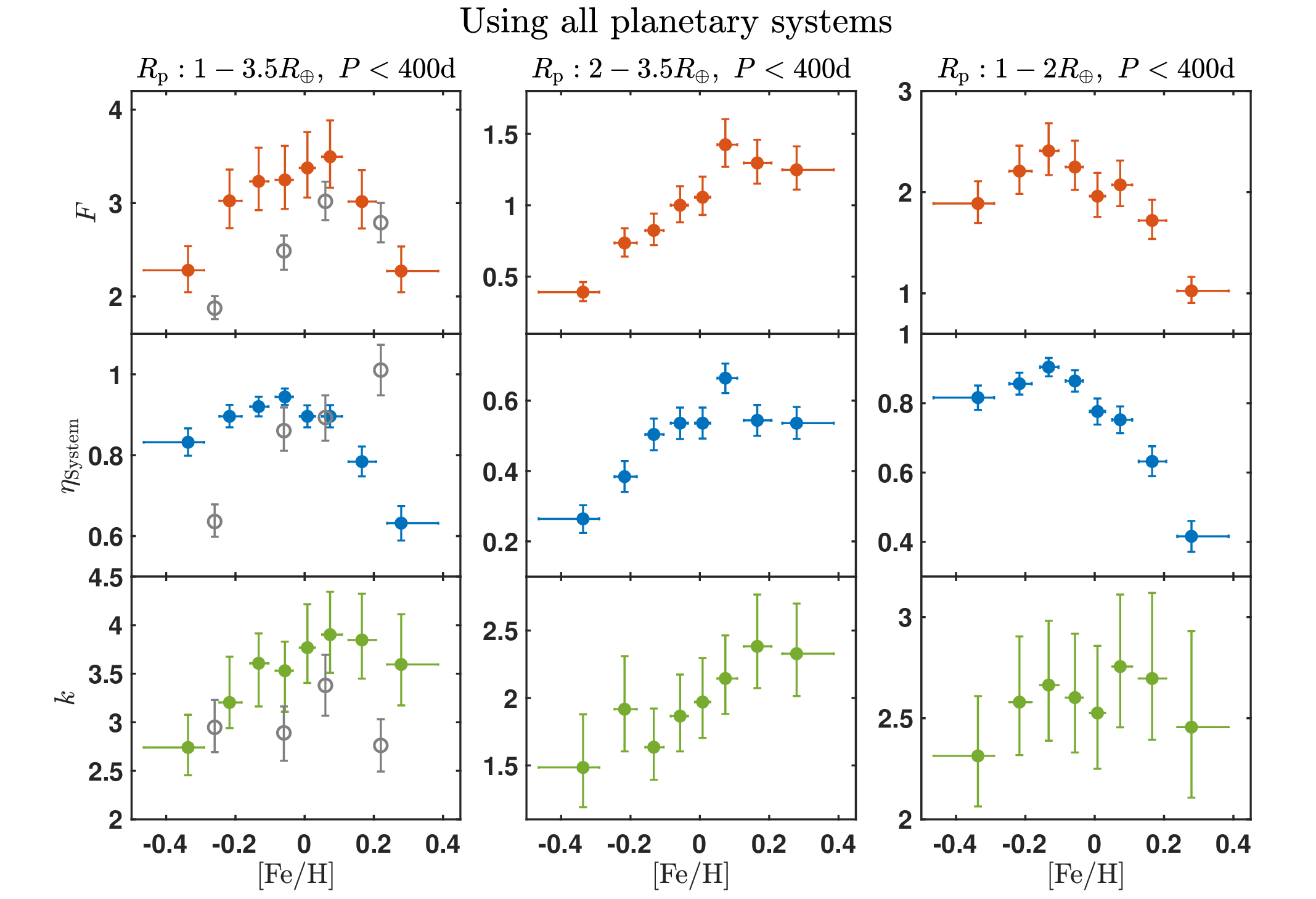}
\caption{The occurrence rate $F$, the fraction of stars hosting planetary system $\eta_{\rm system}$ and the average multiplicity $k$ as a function of stellar metallicity $\rm [Fe/H]$ for Kepler-like planets ($R_{\rm p}: 1-4 R_\oplus$, $P<400$ d, Left panels), planets with radii between $2-4 R_\oplus$ (Middle panels) and planets with radii between $1-2 R_\oplus$ (Right panels) from the synthetic population by Bern planet formation and evolution model at 5 Gyr.
In the Left panels, we also plot the observational results derived from Kepler data \citep{2019ApJ...873....8Z} as grey points and errorbars.
As mentioned before, we are more concerned with trends and thus we renormalize the amplitude of observed occurrence rate tracer of Kepler systems with planets into the average synthetic $\eta_{\rm System}$.
\label{figFKep_FeH_Synthesis}}
\end{figure*}

To explore the correlation between stellar $\rm [Fe/H]$ and the occurrence rates of hot/warm/cold Jupiters, we divide the selected sample into several bins with equal numbers of stars according to their $\rm [Fe/H]$.
Then we count the numbers of hot/warm/cold Jupiters and stars of each bins and obtain the occurrence rates $F = \frac{N_{\rm p}}{N_{\rm S}}$ for different $\rm [Fe/H]$.
To obtain the uncertainties of occurrence rates, we
assume that the numbers of giant planets and stars in each bin obey the Poisson distribution and resample for 1,000 times. 
The uncertainty (1-$\sigma$ interval) of $\rm [Fe/H]$ is set as the range of $50 \pm 34.1$ percentiles of given distribution in each bin.
Figure \ref{figFHJCJWJ_FeH_Synthesis} shows the occurrence rates of hot Jupiters $F_{\rm HJ}$ (top panel), warm Jupiters $F_{\rm WJ}$ (middle panel) and cold Jupiters $F_{\rm CJ}$ (bottom panel) as a function of stellar $\rm [Fe/H]$.
{\color{black} As can be seen, the occurrence rates of hot/warm/cold all seem to increase with stellar $\rm [Fe/H]$.}

To quantify the above trends, referring to previous studies \citep[e.g.][]{2010PASP..122..905J}, we fit the occurrence rates as functions of metallicity with an exponential formula:
\begin{equation}
  F({\rm [Fe/H]}, t) = C \times 10^{\beta \rm [Fe/H]}.
\end{equation}
To obtain the values and uncertainties of $C, \beta$ (hereafter, we denote these coefficients as $X$ for conciseness), we fit them from the number of planets $H$ drawn from a larger sample of $T$ stars using Bayes’ theorem with the similar procedure as described in Sect. 4.2 of \citet{2010PASP..122..905J}.
Specifically, for a given set of $X$, the probability of star $i$ with a planet is $F ({\rm [Fe/H]}_i)$ {\color{black} and the probability of star $j$ without of a planet is $1-F ({\rm [Fe/H]_j})$)}.
Thus, the probability/likelihood of forming $H$ planets from $T$ targets for a given $X$ is:
\begin{equation}
  P \propto \prod \limits_{i}^{H} (F ({\rm [Fe/H]}_i) \times \prod \limits_{j}^{T-H} [1-(F ({\rm [Fe/H]}_j)].
\end{equation}
The marginalized log-likelihood is:
\begin{equation}
\begin{split}
  \log L & = \sum \limits_{i}^{H} \log(F ({\rm [Fe/H]}_i)) \\
  & + \sum \limits_{j}^{T-H} \log[1-F ({\rm [Fe/H]}_j)].
\end{split}
\end{equation}
The best-fits of $X=[C, \ \beta]$ are set as where the maximum $L$ conditioned on the simulated sample and the confidence intervals
(1-$\sigma$: 68.3\%, 2-$\sigma$: 95.4\% and 3-$\sigma$: 99.7\%) are estimated using a spline function.

In Fig. \ref{figFHJCJWJ_FeH_Synthesis}, we show the probability density functions of $X$ for the hot Jupiters, warm Jupiters, and cold Jupiters in the top, middle, and bottom panels, respectively.
As can be seen, all the occurrence rates of hot Jupiters, warm Jupiters, and cold Jupiters are positively correlated ($\beta$ of $1.3^{+0.9}_{-0.6}$, $1.3^{+0.3}_{-0.2}$ and $1.3^{+0.2}_{-0.1}$) with the stellar metallicity with confidence levels (i.e. the probabilities for $\beta>0$) of 96.76\%, 99.99\%, 99.99\%, respectively.
In order to further test the dependence of $F_{\rm HJ}$, $F_{\rm WJ}$ and $F_{\rm CJ}$ on stellar $\rm [Fe/H]$, we also adopt a constant model (i.e. $\beta = 0$) model to fit the synthetic data.
We calculate their Akaike information criterion (AIC).
The resulting AIC differences between the best-fits of the constant model and exponential model are 6.6, 19.9, and 15.4 for $F_{\rm HJ}$, $F_{\rm WJ}$ and $F_{\rm CJ}$ respectively, suggesting the exponential model is confidently preferred.
The above analyses demonstrate that the occurrence rates of the synthetic giant planets are positively-correlated with stellar metallicity.

The dependences of the occurrence rates of giant planets derived from the synthetic population are generally consistent with previous studies based on observational sample \citep[e.g.][]{2001A&A...373.1019S,2004A&A...415.1153S,2005ApJ...622.1102F,
2010PASP..122..905J,2017ApJ...838...25G,2023PNAS..12004179C}.
Specifically, \cite{2010PASP..122..905J} and \cite{2023PNAS..12004179C} find that the occurrence rate of warm/cold Jupiters are positively-correlated with stellar $\rm [Fe/H]$ as $\beta \sim 1.2$ based on RV and Kepler samples, which is in good agreement with our results within 1-$\sigma$ uncertainties.
Observational data also show that the occurrence rate of hot Jupiters exhibits a stronger dependence on $\rm [Fe/H]$ \citep{2017ApJ...838...25G,2023PNAS..12004179C} and the derived $\beta$ is $1.6^{+0.3}_{-0.3}$ after removing the effect of temporal evolution \citep{2023PNAS..12004179C}, which is a bit larger (but statistically
indistinguishable) comparing to $\beta$ from the synthetic sample.
One possible reason is that the synthetic sample does not include the `late-arrived' hot Jupiters that migrated into the close proximity of their host stars long after their parent protoplanetary disks dissipated, which also makes contribution to the observed hot Jupiter populations \citep[e.g. Chen et al. 2025, submitted;][]{2022AJ....164...26H}.
These `late-arrived' hot Jupiters could be delivered via secular chaos in multiple giant planetary systems (Chen et al. 2025, submitted) and thus are expected to have a stronger metallicity dependence since metal-richer stars are more likely to form multiple giant planets \citep[e.g.][]{2013ApJ...767L..24D,2018AJ....155...89P,2020A&A...633A..33S,2023ApJ...949L..21Y}.

\subsection{Neptune-size planets}
\label{sec.frequency.Nep}
In this section, we select planets with radii between $\sim 3.5-6 R_\oplus$ as Neptune-size planets from the synthetic population, yielding a sample of 273 Neptune-size planets (16 hot, 152 warm, and 105 cold).
We further divide the selected sample into several bins according to stellar $\rm [Fe/H]$ and obtain their occurrence rate.
Figure \ref{figFHNCNWN_FeH_Synthesis} shows the occurrence rates of hot/warm/cold Neptune-size planets ($F_{\rm HN}$, $F_{\rm WN}$, and $F_{\rm CN}$) as function of stellar $\rm [Fe/H]$.
Similar to Sect.  4.1, we perform the Bayesian analysis by modelling the planet occurrence rate with the exponential model (Equation 1).
In the right panels, we show the probability density functions of the fitting parameters [$C, \ \beta$].
As can be seen, the occurrence rates of hot Neptunes, warm Neptunes and cold Neptunes are all positively correlated ($\beta$ of $0.5^{+0.5}_{-0.5}$, $0.8^{+0.3}_{-0.2}$ and $0.6^{+0.2}_{-0.3}$) to the stellar metallicity with confidence levels (i.e. the probabilities for $\beta>0$) $\gtrsim 2-3 \sigma$.
That is to say, Neptune-size planets tend to be hosted by metal-richer stars, though this preference is much weaker compared to that of giant planets.

The above results derived from the synthetic Neptune-size planetary population are consistent with those derived from observational data.
Specifically, based on the LAMOST-Kepler data, \cite{2018PNAS..115..266D} found that hot-Neptunes are preferentially hosted by metal-rich stars and their occurrence rates increase with stellar $\rm [Fe/H]$, which was then confirmed
by \cite{2018AJ....155...89P} with the Keck spectra.
By using the LAMOST-Gaia-Kepler kinematic catalogue \citep{2021AJ....162..100C}, \cite{2022AJ....163..249C} found that the fraction of Neptune-size planets with orbital period in the range of $5-100$ day increases with stellar $\rm [Fe/H]$ \citep[$\beta= 0.8^{+0.2}_{-0.4}$;][]{2022AJ....163..249C}.
{\color{black} Recently, based on the homogeneous catalogue
of stellar metallicities from Gaia Data Release 3, \cite{2025AJ....169..117V} found that hottest Neptunes  tend to orbit metal-rich stars.
\cite{2025MNRAS.539.3138D} also reported an enhanced metallicity for our
sample of Neptune desert planets from a homogeneous TESS
sample with HARPS precise radial velocities, consistent with previous observational studies and our synthetic results.}

\begin{figure*}[!t]
\centering
\includegraphics[width=\textwidth]{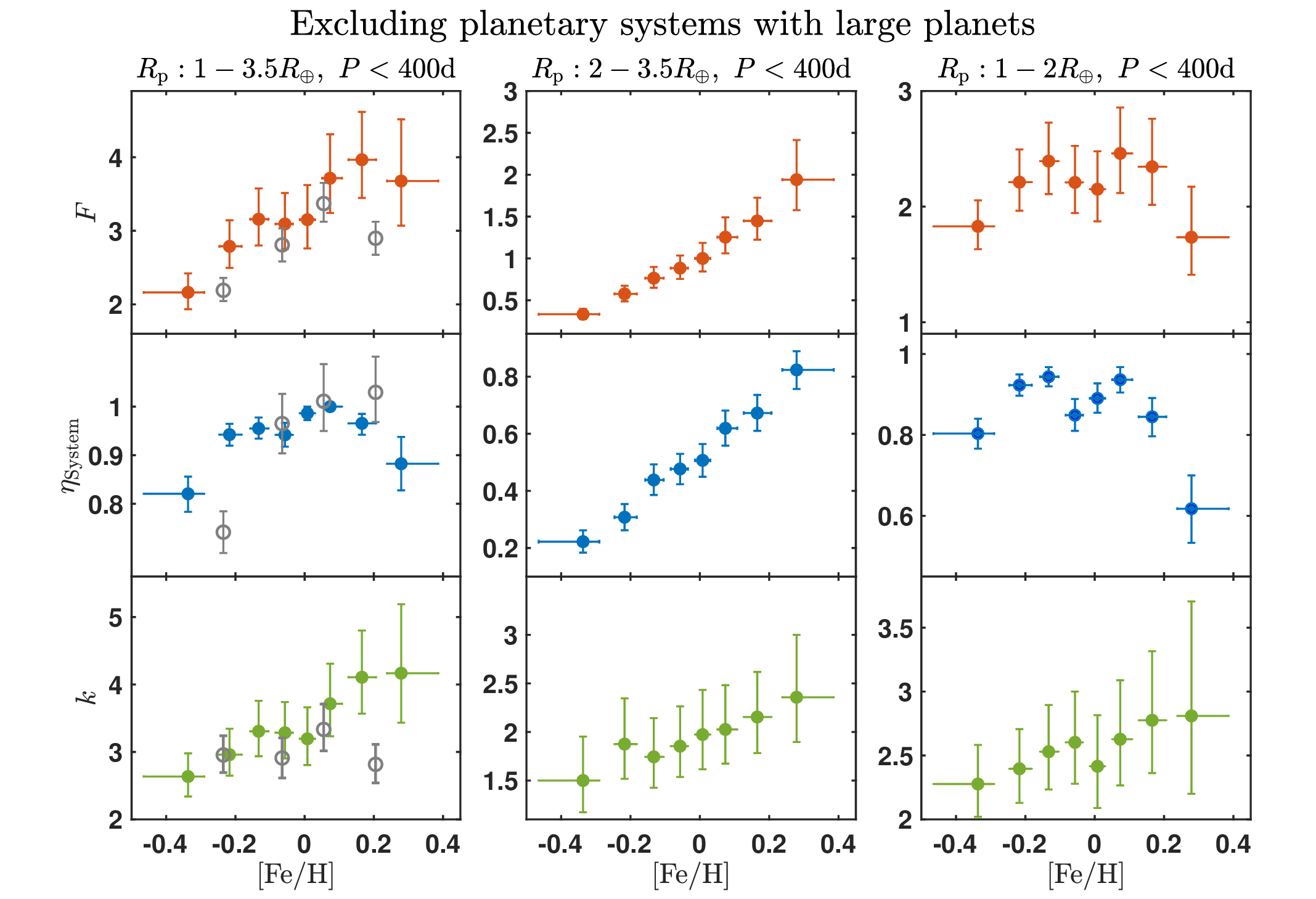}
\caption{Similar to Fig. \ref{figFKep_FeH_Synthesis} but excluding planetary systems including planets with radii $>3.5 R_\oplus$.
\label{figFKep_FeH_SynthesisNonLP}}
\end{figure*}

\subsection{\color{black} Small planets ($1-3.5 R_\oplus$)}
\label{sec.frequency.Kep}
In this section, we investigate the occurrence rate of small planets (i.e. radii within $1-3.5 R_\oplus$ and period less than 400 days) as a function of stellar $\rm [Fe/H]$.
The synthetic population contains 2,948 small planets in 850 systems.
Referring to previous statistical studies on Kepler data \citep[e.g.][]{2018ApJ...860..101Z,2019ApJ...873....8Z,2020AJ....159..164Y}, we further decompose the occurrence rate $F$ into two factors:
the fraction of stars hosting  planetary systems $\eta_{\rm system}$ and the average number of Kepler-planets in one system $k$.

With the similar procedure described in Sect. 4.1, we obtain $F$, $\eta_{\rm system}$ and $k$ for different bins of different $\rm [Fe/H]$.
Then we fit the correlation between $F$, $\eta_{\rm system}$ and $k$ as a function of $\rm [Fe/H]$ with constant model and exponential model, and calculate their difference in AIC score $\rm \Delta AIC \equiv AIC_{\rm Con}-AIC_{\rm Exp}$.
In the left panels of Fig. \ref{figFKep_FeH_Synthesis}, we show the occurrence rate $F$ as well as its two factors $\eta_{\rm system}$ and $k$ as a function of stellar $\rm [Fe/H]$.
As shown in Fig. \ref{figFKep_FeH_Synthesis}, when $\rm [Fe/H] \lesssim$ 0.1 dex, the occurrence rate $F$ as well as the two decomposed factor $\eta_{\rm system}$ and $k$ increase with stellar $\rm [Fe/H]$ with $\rm \Delta AIC$ of 8.4, 3.9 and 12.4, respectively.
In contrast, when $\rm [Fe/H] \gtrsim$ 0.1 dex, $F$, $\eta_{\rm system}$ and $k$ start to decline with $\rm \Delta AIC$ of 10.5, 13.4 and 5.2, respectively.
The trends of first increasing and then decreasing as well as the inflection point at $\sim 0.1$ dex are generally consistent with the observational results of \cite{2019ApJ...873....8Z} derived from LAMOST-Kepler data.

We further divide the small planets into two subsamples: planets with radii $\ge 2 R_\oplus$ and $< 2 R_\oplus$.
As shown in the middle panels of Fig. \ref{figFKep_FeH_Synthesis}, $F$, $\eta_{\rm system}$ and $k$ of planets with radii $\ge 2 R_\oplus$ rise continuously with increasing $\rm [Fe/H]$ ($\rm \Delta AIC$ of 16.3, 13.3 and 7.6) when $\rm [Fe/H] \lesssim 0.1$ dex. 
We also note that $F$, $\eta_{\rm system}$ and $k$ all seem to reach a plateau when $\rm [Fe/H] \gtrsim 0.1$ dex.
For planets with radii $<2 R_\oplus$, the inflection point of $\rm [Fe/H]$ for $F$ and $\eta_{\rm system}$ decreases to $\sim -0.1$ dex.
Besides, $F$ and $\eta_{\rm system}$ decline significantly ($\rm \Delta AIC$ of 9.2 and 12.2) when $\rm [Fe/H] \gtrsim -0.1$ dex, while the average multiplicity $k$ changes mildly (within 1-$\sigma$ error-bar) with $\rm [Fe/H]$.
That is to say, planets with radii $\ge 2 R_\oplus$ are more likely to be hosted by metal-rich stars comparing to planets with radii $< 2 R_\oplus$, which is consistent with previous statistical studies based on observation data \citep[e.g,][]{2022AJ....163..249C}.

\begin{figure}[!t]
\centering
\includegraphics[width=0.95\linewidth]{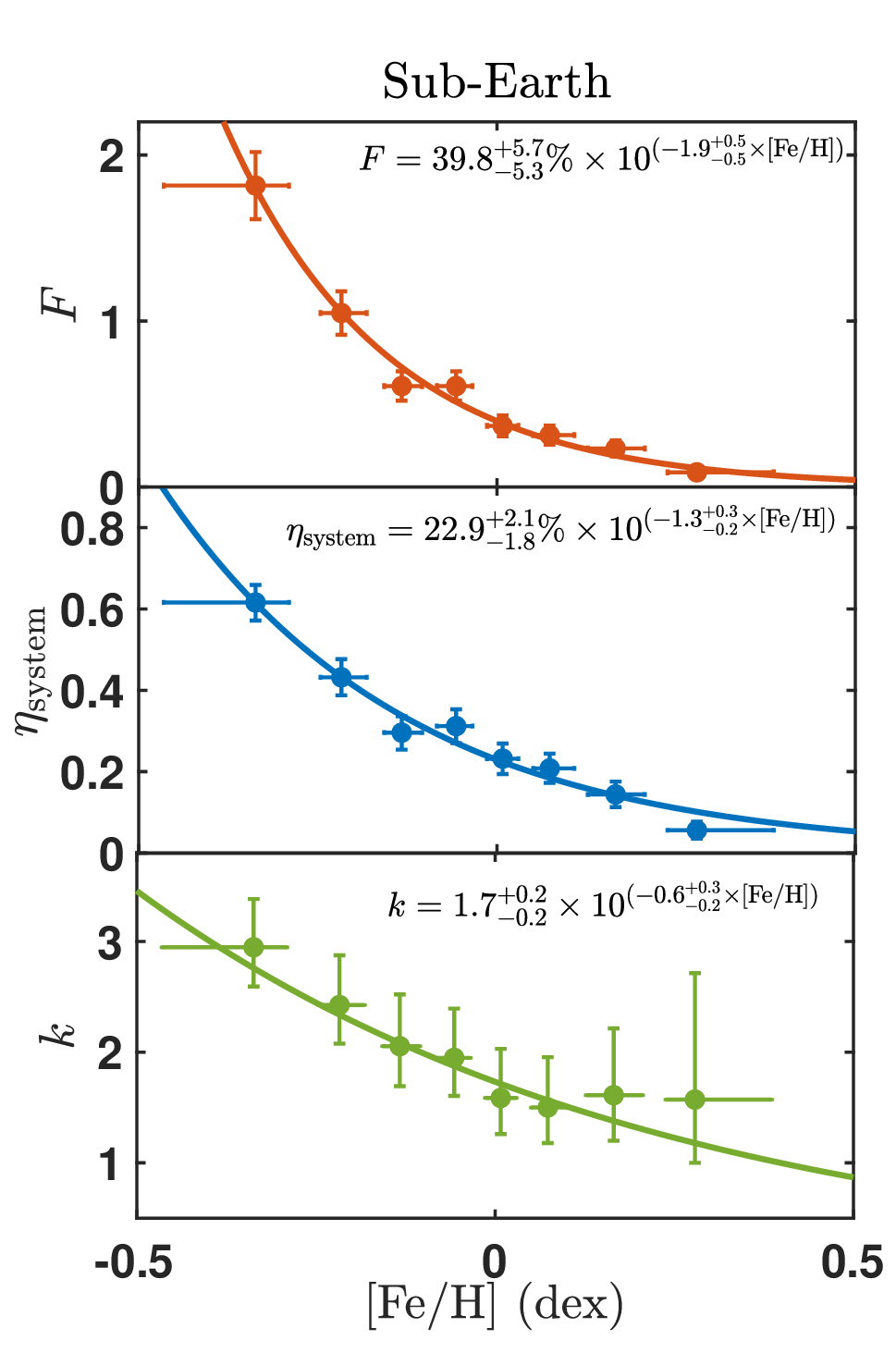}
\caption{The occurrence rate $F$ (Top), the fraction of stars hosting planetary system $\eta_{\rm system}$ (Middle) and the average multiplicity $k$ (Bottom) as a function of stellar metallicity $\rm [Fe/H]$ for the sub-Earths from the synthetic population by Bern planet formation and evolution model at 5 Gyr.
\label{figFSubEarth_FeH_Synthesis}}
\end{figure}

One potential explanation for the decline of $F$ at $\rm [Fe/H] \gtrsim 0.1$ dex is that metal-richer stars host more giant planets and Neptunes.
Their perturbation can drive dynamical instabilities \citep[e.g.][]{2007ApJ...666..423Z} and cause merge/ejection of small planets \citep[e.g.][]{2015ApJ...807...44P,2017AJ....153...42L}, reducing the number of planets or even the number of systems including small planets. 
To test whether this explanation is feasible, we exclude planetary systems with large planets with radii $>3.5 R_\oplus$ (regardless of whether period larger or less than 400 days) and divide the rest 624 planetary systems into eight bins with the same intervals of $\rm [Fe/H]$ as before.
As shown in the bottom panels of Fig. \ref{figFKep_FeH_SynthesisNonLP}, after eliminating the influence of large planets, the decrease of average multiplicity $k$ at $\rm [Fe/H] \gtrsim 0.1$ disappears for small planets as well as the two subsamples (with larger $\rm \Delta AIC$ of 18.1, 17.6 and 12.1).
Furthermore, the declines of $F$ and $\eta_{\rm system}$ at super-solar metallicity disappear for the planets with radii $\ge 2 R_\oplus$ and become much weaker for those with radii $<2 R_\oplus$.
These changes are in good agreement with the prediction of the proposed explanation and thus support that the perturbation by larger planets is an important mechanism causing the decline of the occurrence rate at super-solar metallicity.
Nevertheless, the decreasing of and $F$ and $\eta_{\rm system}$ of planets with radii $< 2 R_\oplus$ at $\rm [Fe/H] \gtrsim 0.1$ dex maintains, may indicating protoplanetary disks with super-solar metallicities have too many solids to form planets with radii $< 2 R_\oplus$.

\subsection{Sub-Earths}
\label{sec.frequency.subEar}
Sub-Earths are generally referred to planets smaller than our Earth, which are most difficult to be detected due to the weakest signals.
For observation, recently,
Kepler data reveals an inflection in the occurrence rate of planets around $\sim 1 R_\oplus$, suggesting that sub-Earths represent a unique population which are not an extension of super-Earths \citep{2019AJ....158..109H,2020ApJ...891...12N,2021AJ....161..201Q}.
However, the current sample of sub-Earth is not large enough to constrain their occurrence rate as a function of metallicity reliably.

Based on the synthetic population generated by the Bern model, we obtain a sample of 635 sub-Earths with radii between $0.3-1 R_\oplus$ and orbital period $\le 400$ days.
investigate the correlations between their occurrence rate and stellar $\rm [Fe/H]$.
We divide the stellar sample into eight bins according to $\rm [Fe/H]$ and obtain the occurrence rate $F$, the fraction of stars hosting sub-Earths $\eta_{\rm system}$ and the average multiplicity $k$ of sub-Earths for different bins of different $\rm [Fe/H]$.
Figure \ref{figFSubEarth_FeH_Synthesis} shows the evolution of $F$, $\eta_{\rm system}$ and $k$ with stellar $\rm [Fe/H]$ for sub-Earths. As can be seen, $F$ as well as $\eta_{\rm system}$ and $k$ of the sub-Earths all decrease significantly with increasing $\rm [Fe/H]$.
To quantify the above trends, we make fitting and analyses with the same procedure described in Sect. 4.1. The best-fits of preferred models are plotted as solid lines in Fig. \ref{figFSubEarth_FeH_Synthesis}. The exponential models are preferred for $F$, $\eta$ and $k$ of sub-Earths compared to the constant model with $\rm \Delta AIC$ of 33.7, 26.4 and 15.3, respectively.
$F$, $\eta$ and $k$ are negatively correlated with stellar $\rm [Fe/H]$ with confidence levels (i.e. $\beta<0$) of 99.99\%, 99.98\% and 95.09\%, respectively.
The above analyses suggest that sub-Earths tend to be hosted by metal-poor stars with confidence level $\gtrsim 2-3 \sigma$. 
This result is expected because  around metal-rich stars the amount of silicates/Fe is more likely to be sufficiently high that planets more massive/larger than sub-Earths are preferentially formed.

\section{Radius valley morphology as functions of stellar metallicity}
\label{sec.valley}

One of the most interesting discoveries in exoplanet science in recent years is the radius valley, a dip at $\sim 1.9 R_\oplus$ in the radius distribution as revealed by high-precision sample from Kepler \citep[e.g.][]{2017AJ....154..109F,2018MNRAS.479.4786V}.
Plenty of previous statistical studies have explored the dependence of the radius valley on planetary and stellar properties, providing crucial insights into the origin of the radius valley \citep[e.g.][]{2017AJ....154..109F,2018MNRAS.479.4786V,2018MNRAS.480.2206O,2020AJ....160..108B,2022AJ....163..249C}.
The radius valley had been theoretically predicted to exist \citep{2013ApJ...775..105O,2013ApJ...776....2L,2014ApJ...795...65J} before it was discovered in the observational data.
In a previous paper \citep{2024NatAs...8..463B}, we find that the synthetic \texttt{NG76Longshot} population can reproduce the radius distributions with a similar position of the valley centre $\sim 1.9 R_\oplus$ \citep[e.g.][]{2013ApJ...775..105O,2017ApJ...847...29O,2017AJ....154..109F}, the radius cliff at $\sim 3 R_\oplus$ \citep[e.g.][]{2019ApJ...887L..33K} and the anti-correlation on orbital period \citep{2018MNRAS.479.4786V,2019ApJ...875...29M} revealed from observational data. 
In this subsection, we further investigate the morphology of the radius valley as a function of stellar metallicity quantitatively.
This is an interesting theoretical prediction that should be tested with observational data.

\begin{figure}[!t]
\centering
\includegraphics[width=1.0\linewidth]{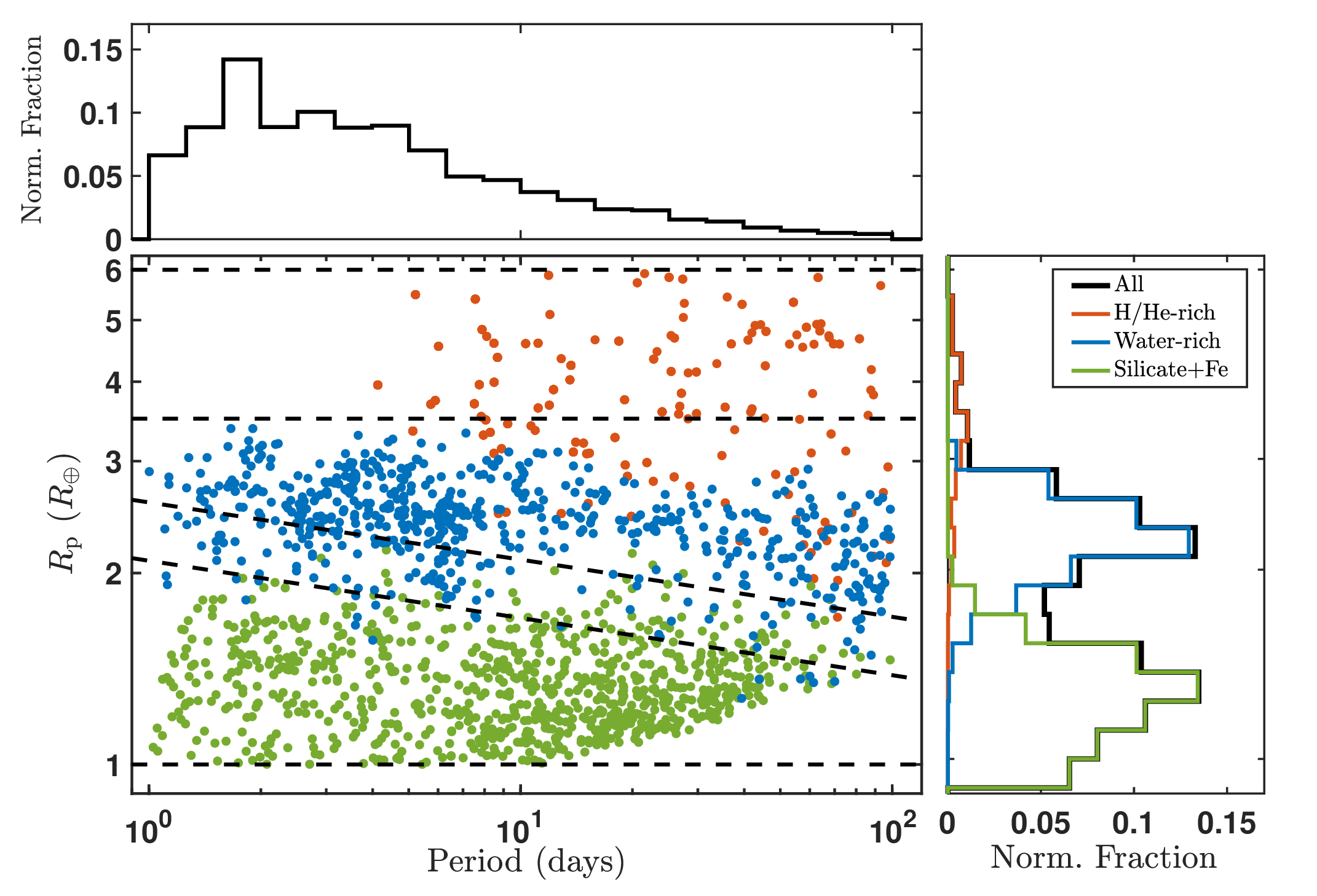}
\caption{\color{black} The period-radius diagram of planets selected from the synthetic population at 2 Gyr after applying the detection selection similar to PAST \uppercase\expandafter{\romannumeral3} \citep{2022AJ....163..249C}. 
Planets with different compositions are plotted in different colours:
red for H/He-rich envelopes, blue for water-rich envelopes, and green for purely rocky (Silicate$+$Fe) planets.
The dashed lines represent the boundaries of planet radii between different sub-samples: Neptune-size planets, sub-Neptune, valley planet, super-Earth).
Histograms of planetary radius and orbital period are shown in the right panel and the top panel, respectively.
\label{figRadiusValley}}
\end{figure}

\begin{figure*}[!t]
\centering
\includegraphics[width=0.8\textwidth]{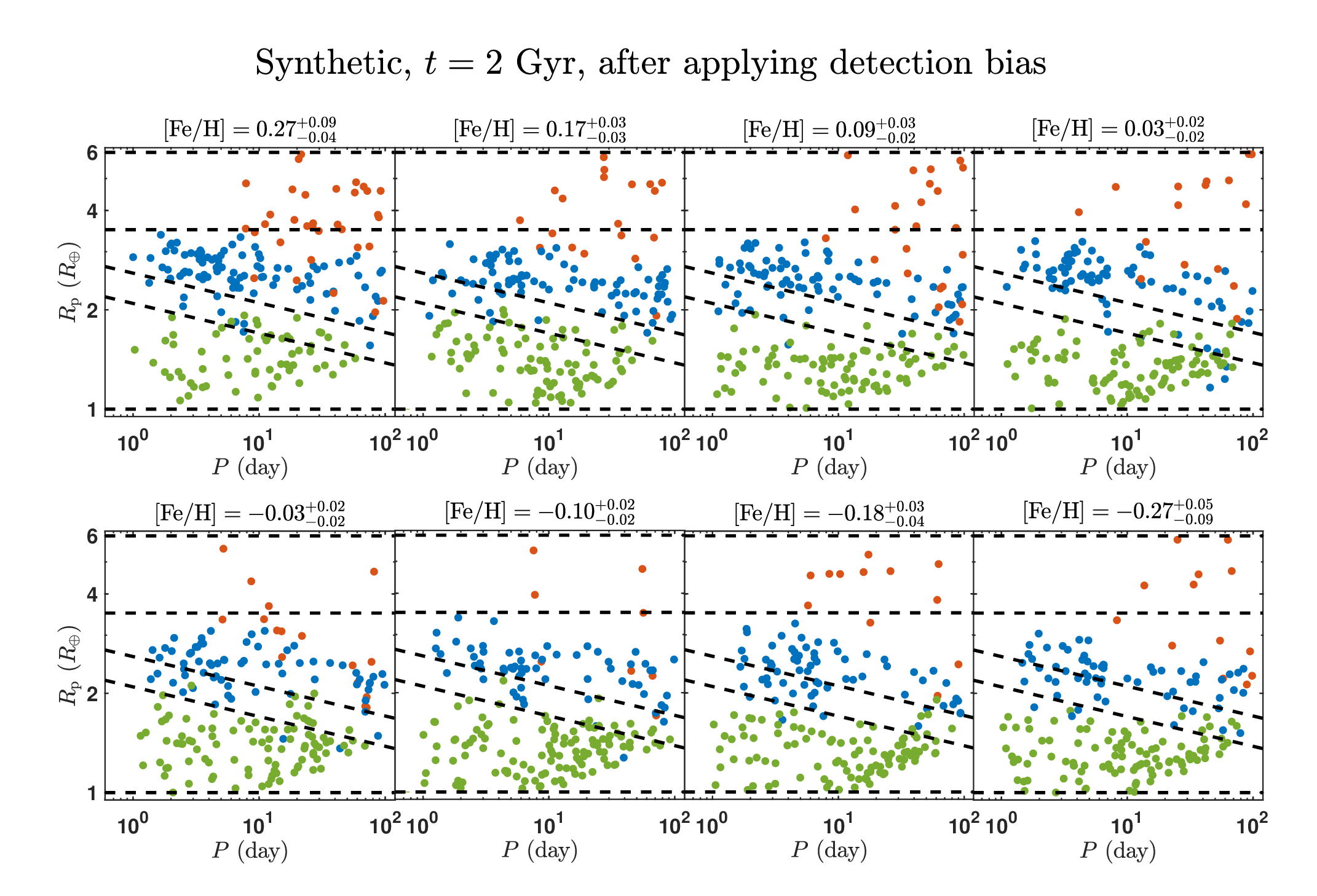}
\caption{\color{black} Orbital period-radius diagram of planets with different metallicity $\rm [Fe/H]$ from the synthetic population at 2 Gyr after applying the detection bias similar to PAST \uppercase\expandafter{\romannumeral3}. 
\label{figRadiusValleyFeH}}
\end{figure*}

\begin{figure*}[!t]
\centering
\includegraphics[width=0.8\textwidth]{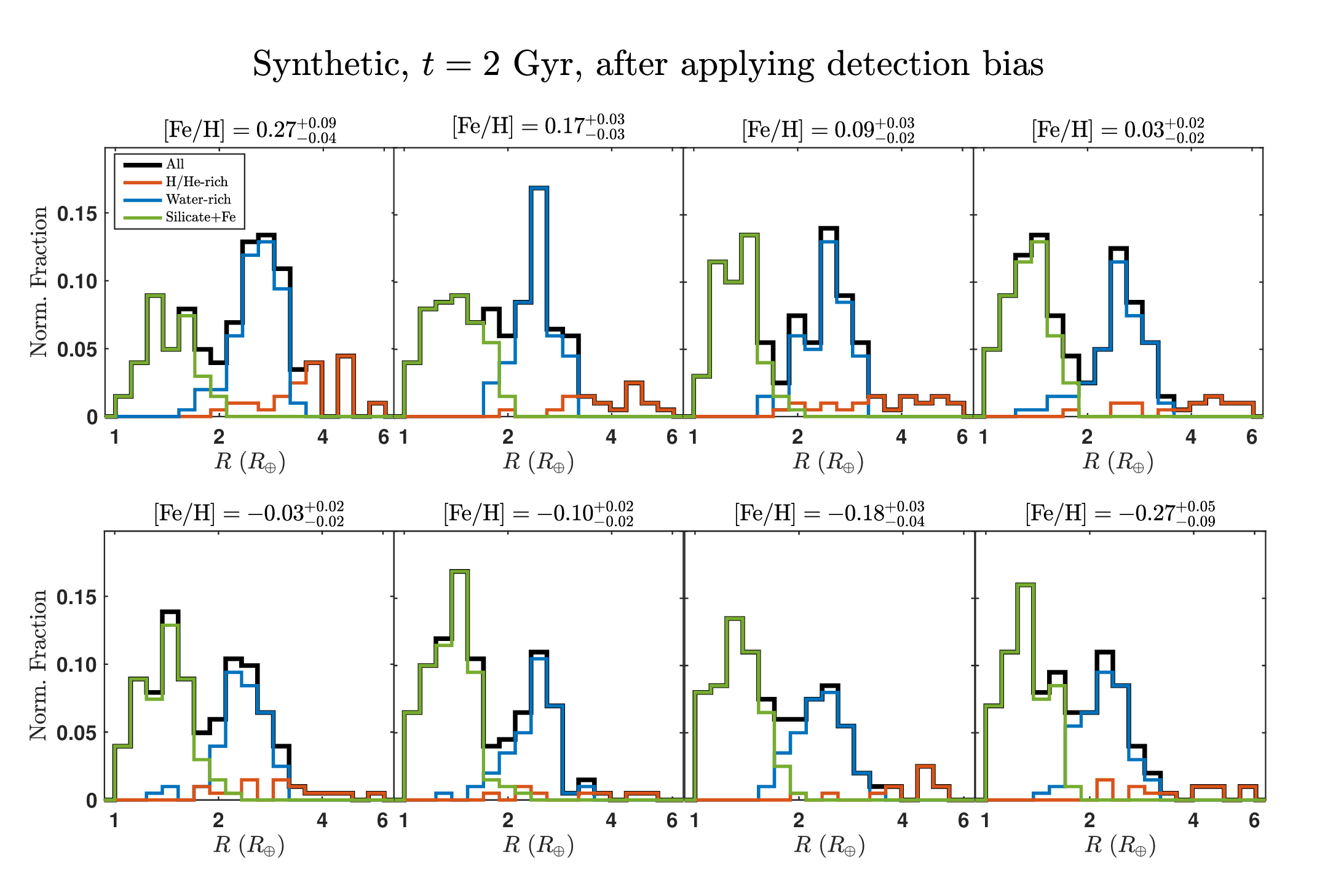}
\caption{\color{black} Radius distribution of planets with different metallicity $\rm [Fe/H]$ from the synthetic population after applying a detection selection similar to PAST \uppercase\expandafter{\romannumeral3}.
\label{figRadiushistFeH}}
\end{figure*}

We initialize our planetary sample from the synthetic planetary population at 2 Gyr (the typical age of planetary sample of \cite{2022AJ....163..249C} after controlling parameters) and select planets close to the radius valley (i.e. radius between $1-6 R_\oplus$) with orbital period between $1-100$ days.
To apply the detection biases, we simulate the Kepler mission completeness using the Kepler Observes Bern Exoplanets (KOBE) program \citep{2021A&A...656A..74M}.
Furthermore, we also exclude planets with detection efficiency less than 50\% since most of Kepler planets in \citet{2022AJ....163..249C} lie above the 50\% detection efficiency limit (see Fig. B3 of their paper).
Figure \ref{figRadiusValley} displays the planetary radius-period diagram and planetary radius histogram. 
As can be seen, the synthetic sample shows an obvious bimodal radius distribution (with $p-$value $<0.003$ from maximizes Hartigan's dip statistic).

{\color{black} The radius valley are found to be located at $R^{\rm Valley}_{\rm p0} \sim 1.9 \pm 0.2 R_\oplus$ \citep[e.g.][]{2017ApJ...847...29O,2017AJ....154..109F,2018ApJ...853..163J} and to be dependent on stellar mass $M_*$ \citep{2020AJ....160..108B} and planetary orbital period $P$ \citep{2018MNRAS.479.4786V}, which can be characterized as:
\begin{equation}
    R^{\rm valley}_{\rm p} = R^{\rm valley}_{\rm p0}
    \left(\frac{M_*}{M_\odot}\right)^h
    \left(\frac{P}{\rm 10 days}\right)^g,
\end{equation}
where $h$ and $g$ are are the corresponding slopes, which are adopted as $0.21^{+0.06}_{-0.07}$ and $-0.09^{+0.02}_{-0.03}$ \citep{2023MNRAS.519.4056H,2024MNRAS.531.3698H}, respectively.
For our synthetic sample, $M_*=1 M_\odot$ and thus the upper and lower boundaries of radius valley are set as $R^{\rm Valley}_{\rm upper} = 2.1 R_{\oplus} \left(\frac{P}{\rm 10 days}\right)^{-0.09}$ and $R^{\rm Valley}_{\rm lower} = 1.7 R_{\oplus} \left(\frac{P}{\rm 10 days}\right)^{-0.09}$, respectively.
Thus, we divide the planetary sample into four sub-samples according to their radius as follows: 
\begin{enumerate}
    \item Valley planet (VP):
    $R^{\rm Valley}_{\rm lower}  \le R_{\rm p} \le R^{\rm Valley}_{\rm upper}$;
    \item Super-Earth (SE): $1 R_{\oplus} \le R_{\rm p}  <R^{\rm Valley}_{\rm lower}$;
    \item Sub-Neptune (SN): $R^{\rm Valley}_{\rm upper}< R_{\rm p} \le 3.5 R_{\oplus}$;
    \item Neptune-size planet (NP): $3.5 R_{\oplus}< R_{\rm p} \le 6 R_{\oplus}$.
\end{enumerate}}
Moreover, using the composition data provided by the synthetic population, the planetary sample can be classified into three categories: pure rocky planets (silicate$+$Fe) with no envelope, planets with water-rich envelope (mass fraction of water in the envelope $\ge 0.9$), and planets with H/He-rich envelope (mass fraction of water $<0.9$). As can be seen in Fig. \ref{figRadiusValley}, for the synthetic sample, Neptune-like planets all have H/He-rich envelopes. Super-Earths are mostly purely rocky which mainly ($\sim 75\%$) formed in-situ by a giant impact stage (Class I system architecture and formation pathway in the classification of \citealt{2023EPJP..138..181E}) or lost their gaseous envelope completely (more relevant at short orbits and larger super-Earths), while sub-Neptunes are mainly water-rich planets that formed ex-situ and migrated substantially to their present-day location (Class II in \citealt{2023EPJP..138..181E}).


To investigate the morphology of radius valley as a function of stellar metallicity, we divide the planetary sample into eight bins with same size according to $\rm [Fe/H]$.
Figure \ref{figRadiusValleyFeH} and \ref{figRadiushistFeH} display the radius-period diagrams and radius histograms for planets with different metallicities.
As can be seen, with the decrease of $\rm [Fe/H]$, the radius valley becomes less prominent.
Also, the number (fraction) of super-Earths grows, while the numbers (fractions) of sub-Neptunes and Neptunes-like planets decline. 
Moreover, the radii of sub-Neptunes seem to shrink with decreasing $\rm [Fe/H]$.
Very interestingly, such trends are consistent with the observational results derived from the CKS sample \citep{2018MNRAS.480.2206O} and LAMOST-Gaia-Kepler sample \citep{2022AJ....163..249C}.

\begin{figure}[!t]
\centering
\includegraphics[width=0.5\textwidth]{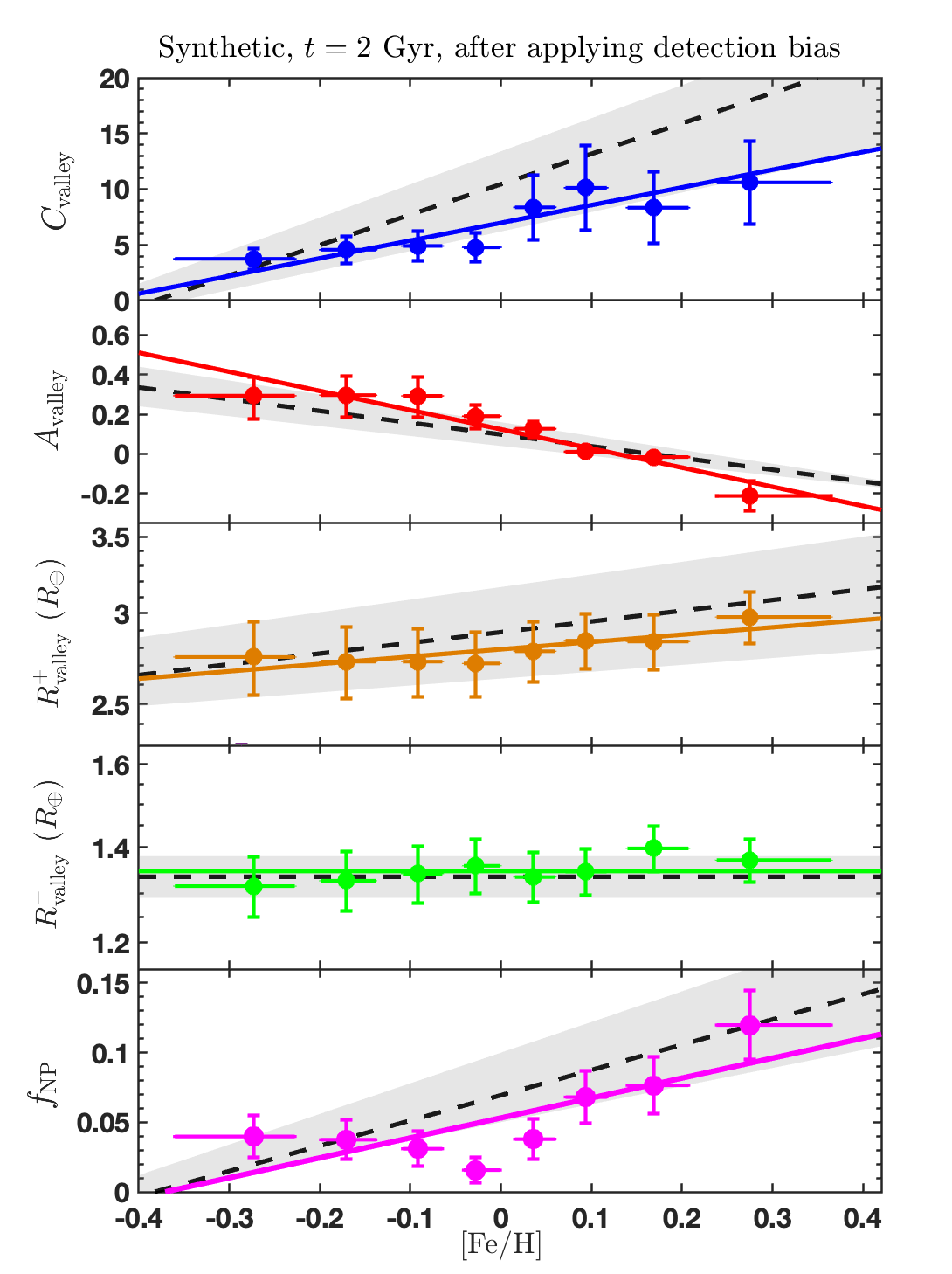}
\caption{\color{black} The five metrics to characterize the radius valley morphology (i.e. $C_{\rm valley}$, $A_{\rm valley}$, ${R}^{+}_{\rm valley}$, ${R}^{-}_{\rm valley}$, and $f_{\rm NP}$) as functions of  $\rm [Fe/H]$ for synthetic population at 2 Gyr after applying the detection bias.
The solid lines denote the best fits derived from synthetic population.
The best fits and 1-$\sigma$ interval derived from the 
selected observational sample of LAMOST-Gaia-Kepler catalogue \citep{2021AJ....162..100C,2022AJ....163..249C} are plotted as dashed black lines and grey regions.
\label{figRadiusmetricFeH}}
\end{figure}

To evaluate the above trends quantitatively, we adopt the five morphology metrics of radius valley defined in \cite{2022AJ....163..249C}, which are described as follows:
\begin{enumerate}
    \item The contrast of radius valley $C_{\rm valley}$, defined as the number ratio of super-Earths ($N_{\rm SE}$) plus sub-Neptunes ($N_{\rm SN}$) to the valley planets ($N_{\rm VP}$), i.e.
    \begin{equation}
    C_{\rm valley} = \frac{N_{\rm SE}+N_{\rm SN}}{N_{\rm VP}}.
    \label{eqDV}
    \end{equation}
    \item  The asymmetry on the two sides of the valley defined as the number ratio (in logarithm) of super-Earths to sub-Neptunes, i.e. \begin{equation}
    A_{\rm valley} = \log_{10} \left( \frac{N_{\rm SE}}{N_{\rm SN}} \right).
    \label{eqREN}  
    \end{equation} 
    \item The average radius of planets with size larger than Valley planets, i.e. 
    \begin{equation}
    {R}_{\rm valley}^{+} = \frac{1}{N_{\rm SN}+N_{\rm NP}}\sum^{N_{\rm SN}+N_{\rm NP}}_{i}{R_{\rm i}}, 
    \label{eqRNSE}  
    \end{equation}
    where {\color{black} $R^{\rm Valley}_{\rm upper} <R_{\rm i}<6R_\oplus$.}
    \item The average radius of planets with size smaller than Valley planets, i.e. 
    \begin{equation}
    {R}_{\rm valley}^{-} = \frac{1}{N_{\rm SE}}\sum^{N_{\rm SE}}_{i}{R_{\rm i}}, 
    \label{eqRSE}  
    \end{equation}
    where {\color{black} $1.0R_\oplus <R_{\rm i}<R^{\rm Valley}_{\rm lower}$.}
    \item The number fraction of Neptune-size planets in the whole planet sample , i.e.
    \begin{equation}
    f_{\rm NP} = \frac{N_{\rm NP}}{N_{\rm SE}+N_{\rm VP}+N_{\rm SN}+N_{\rm NP}}.
    \label{eqfNep}  
    \end{equation}
\end{enumerate}

\begin{figure*}[!t]
\centering
\includegraphics[width=0.9\textwidth]{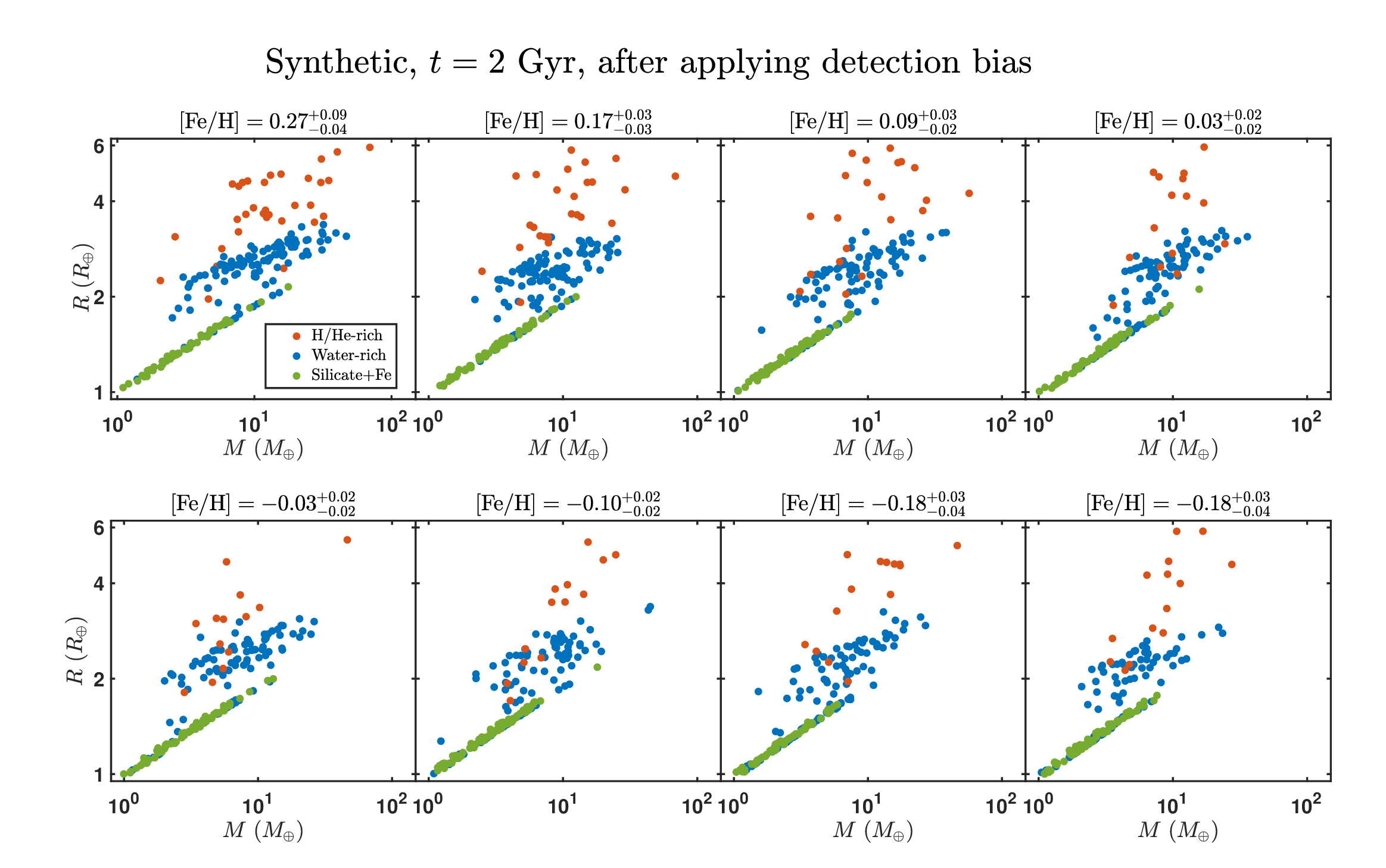}
\caption{\color{black} The mass-radius distribution of planets with different metallicity $\rm [Fe/H]$ from the synthetic population after applying  the detection selection similar to PAST \uppercase\expandafter{\romannumeral3}.
\label{figRadiusmassFeH}}
\end{figure*}

Then we calculate these metrics for the eight bins of different $\rm [Fe/H]$ (as shown in Fig. \ref{figRadiusmetricFeH}).
The uncertainties of $C_{\rm valley}$, $A_{\rm valley}$ and $f_{\rm NP}$ are derived using error propagation based on the Poisson errors in the planet numbers.
The uncertainties of ${R}_{\rm valley}^{+}$ and ${R}_{\rm valley}^{-}$ are calculated using a bootstrapping method.

{\color{black} We also fit the data points using two models (i.e. constant model and linear model) and evaluated these two models by calculating the difference in AIC score ($\rm \Delta AIC$).
The fitting results are as follows:
\begin{enumerate}
    \item For $C_{\rm valley}$, the linear increasing model is preferred over the constant model with $\rm \Delta  AIC=11.7$.
    The best-fit is 
    \begin{equation}
    C_{\rm valley}= 15.9^{+3.4}_{-4.6} \times {\rm [Fe/H]} + 7.0^{+1.6}_{-1.1}.
    \label{eqC_Fe}
    \end{equation}
    
    \item For $A_{\rm valley}$, the linear decay model is preferred over the constant model with $\rm \Delta  AIC=17.9$. 
    The best-fit is 
    \begin{equation}
    A_{\rm valley}= -1.0^{+0.2}_{-0.2} \times {\rm [Fe/H]}+ 0.12^{+0.01}_{-0.01}.
    \label{eqA_Fe}
    \end{equation}
    
    \item For ${R}^{+}_{\rm valley}$, the linear increasing model is preferred over the constant model with $\rm \Delta  AIC=10.5$.
    The best-fit is 
    \begin{equation}
    \log_{10}\left(\frac{{R}^{+}_{\rm valley}}{R_\oplus}\right)= 0.06^{+0.02}_{-0.02} \times {\rm [Fe/H]} + 0.45^{+0.03}_{-0.02}.
    \label{eqR+_Fe}
    \end{equation}
    
    \item $R^{-}_{\rm valley}$ is indistinguishable from having no dependence on $\rm [Fe/H]$ as the fitting linear index ($0.03 \pm 0.04$) is consistent with zero within 1-$\sigma$.
    The best-fit is 
    \begin{equation}
        {R}^{-}_{\rm valley}= 1.35^{+0.02}_{-0.02} R_\oplus.
    \label{eqR-_Fe}
    \end{equation}
     
    \item For $f_{\rm NP}$, the linear increasing model is preferred over the constant model with $\rm \Delta  AIC=45.6$. 
    The best-fit is 
    \begin{equation}
    f_{\rm NP} = 0.15^{+0.04}_{-0.04} \times {\rm [Fe/H]}+0.05^{+0.02}_{-0.01}.
    \label{eqfNp_Fe}
    \end{equation}
\end{enumerate}}
The fitting results of simulation data confirm the apparent trends seen in Fig. \ref{figRadiusmetricFeH} and suggest that the radius valley deepens with increasing $\rm [Fe/H]$.
Furthermore, metal-rich stars host more (and larger) sub-Neptunes and Neptune-size planets.

Such trends are expected.
As shown in Equation (1), proto-planetary disks with higher $\rm [Fe/H]$ exhibit higher solid surface densities. 
Consequently, as $\rm [Fe/H]$ increases, the planets grow faster and attain larger masses.
More planets can reach the mass required for significant inward migration before the disk dissipates \citep{2023EPJP..138..181E}, leading to higher occurrence rates of Neptunes ($f_{\rm NP}$) and sub-Neptunes in the inner systems. 
Moreover, massive planets have deeper gravitational potentials, allowing them to retain more of their envelopes after photo-evaporation and have larger radii ($R^{+}_{\rm valley}$), as the mass-loss rate follows $\dot{M}/M \propto R^3/M^2 \simeq 1/M$ under energy-limited escape \citep{2020A&A...638A..52M}.
In contrast, super-Earths are mainly rocky planets that formed in-situ, with lower requirements for formation speed and mass. 
As a result, the occurrence rate and average size of super-Earths ($R^{-}_{\rm valley}$) show no significant dependence on $\rm [Fe/H]$. 
This also explains why $A_{\rm valley}$ exhibits a positive trend, driven by the increased occurrence of sub-Neptunes rather than a decrease in super-Earths (see also the right panels of Fig.  \ref{figFKep_FeH_Synthesis}).
For $C_{\rm valley}$, as shown in Fig. \ref{figRadiusmassFeH}, the lowest-mass water-rich planets fill the valley in radius space at low [Fe/H], explaining the increasing trend of $C_{\rm valley}$ with $\rm [Fe/H]$. To explain this, it is insightful to remember that on a distance-mass diagram, inward motion occurs if the migration rate is shorter than the solid accretion rate (gas accretion is not efficient at this mass range yet). Thus, for low [Fe/H] and reduced accretion, orbital migration occurs at lower planet mass, which allows for low-mass water-worlds to reach the inner regions of the system. The outcome is similar to that around low-mass stars where less solid material is available due to less-massive expected disks \citep{2022Sci...377.1211L,2024arXiv241116879B,2024A&A...686L...9V}.

In Fig. \ref{figRadiusmetricFeH}, we also plot the best-fits and 1-$\sigma$ intervals of the observational results derived from the LAMOST-Gaia-Kepler sample \citep{2022AJ....163..249C} as dashed black lines and grey regions.
As can be seen, for all five metrics, the trends with $\rm [Fe/H]$ in the synthetic population are quantitatively consistent with those of \cite{2022AJ....163..249C} within $\sim 1$-$2 \sigma$ error-bars, demonstrating that the synthetic population can reproduce very well the observed dependence of the radius valley morphology on host star metallicity.
This quantitative, and not merely qualitative excellent match is a non-trivial result as the model was not optimized or otherwise conditioned to reproduce these observations.
In contrast, the entire formation and evolution that leads to this outcome is governed by the physical processes included in the model.

\section{The distributions of planets as a function of orbital periods}
\label{sec.distribution_radii}
In this section, using again the synthetic \texttt{NG76Longshot} population generated by the Bern model, we explore
how the planet population depends on host star metallicity as a function of orbital periods. 
We initialized the planetary sample based on the synthetic population at 5 Gyr and selected planets with an orbital period $\le 400$ days.

\begin{figure}[!t]
\centering
\includegraphics[width=0.95\linewidth]{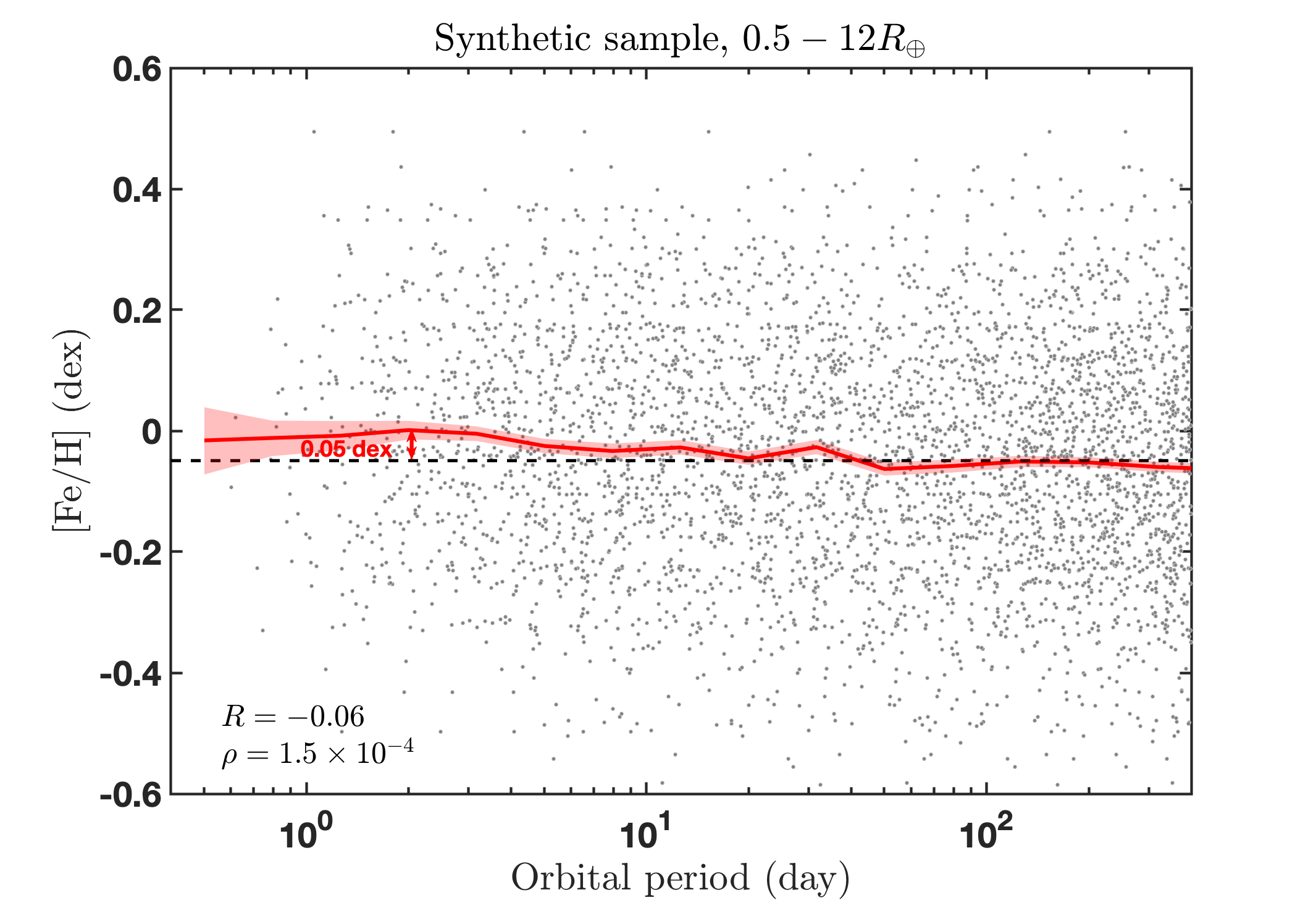}
\caption{Host star metallicities vs. planet orbital period (grey dots) from the synthetic sample. 
The red line denotes the kernel regression of the mean metallicity of the planet population (From Equation \ref{eqmeanFeH}).
The shaded red area shows the $68\%$ (1-$\sigma$) confidence interval on the mean from bootstrapping.
The black dashed line represent the mean metallicity ($-0.05 \pm 0.04$ dex) for stars hosting planets with period $>10$ days.
In the bottom-left corner, we print the Pearson coefficient $R$ between host star metallicity and the orbital periods as well as the correspond $p-$value.
\label{figPeriodFeHALL}}
\end{figure}

Figure \ref{figPeriodFeHALL} displays the metallicities of planet host stars as a function of the orbital period.
To evaluate the correlation between the metallicity and orbital periods, we first adopt the Pearson test and the resulted Pearson coefficient is -0.06 with a $p-$value of $1.5 \times 10^{-4}$ that the two properties are uncorrelated.
We also perform the KS test between the metallicity of stars hosting planets with periods interior and exterior to 10 days.
The resulting $p-$value is 0.015, suggesting that the two subsamples are not drawn from the same distribution.
The above tests demonstrate that the metallicity and orbital periods have a weak but statistically significant anti-correlation, with highest average $\rm [Fe/H]$ at a small distance.
Referring to \cite{2016AJ....152..187M}, we also calculate the kernel regression of the mean metallicity of planet population at an orbital period $P$ with the following formula:
\begin{equation}
    \overline{\rm [Fe/H]} (P) = \frac{\sum^{n}_{i=0}{{\rm [Fe/H]}_{i} f_{{\rm occ},i} K(\log(P/P_i), \sigma)}}{\sum^{n}_{i=0}{f_{occ,i} K(\log(P/P_i), \sigma)}},
    \label{eqmeanFeH}
\end{equation}
where ${\rm [Fe/H]}_{i}$ and $P_i$ are the metallicity and orbital period for each planet in the synthetic sample. 
$f_{\rm occ}$ is the the ratio of the number of planets over the number of stars where planets are detectable, multiplied by the geometric transit probability, which is used to eliminate the effect of detection bias.
We set $f_{\rm occ}=1$ as the synthetic population aim to reveal the intrinsic planet distribution.
Here we adopt the same log-normal kernel with $\sigma = 0.29$ as \cite{2016AJ....152..187M}:
\begin{equation}
    K(\log P, \sigma) = \frac{1}{\sqrt{2\pi}} e^{-0.5 (\log P/ \sigma)^2}.
\end{equation}
The uncertainties of $\overline{\rm [Fe/H]}$ are calculated using a bootstrapping method.
As shown in Fig. \ref{figPeriodFeHALL}, the mean metallicity (solid red line) and 1-$\sigma$ interval (red region) generally decreases with the increase of orbital periods.
We then calculate the mean average metallicities for planet populations with orbital periods interior and exterior to 10 days using Eq. \ref{eqmeanFeH} and their difference,
\begin{equation}
    \Delta \rm [Fe/H] = \overline{\rm [Fe/H]} (P<10 \,  {\rm days}) - \overline{\rm [Fe/H]} (P>10 \, {\rm days})
\end{equation}
is $0.05^{+0.01}_{-0.01}$. 
Out of 10,000 sets of resample data, $\Delta \rm [Fe/H]>0$ for 9,999 times, suggesting that inner planets tend to be hosted by metal-richer stars.

\begin{figure}[!t]
\centering
\includegraphics[width=0.9\linewidth]{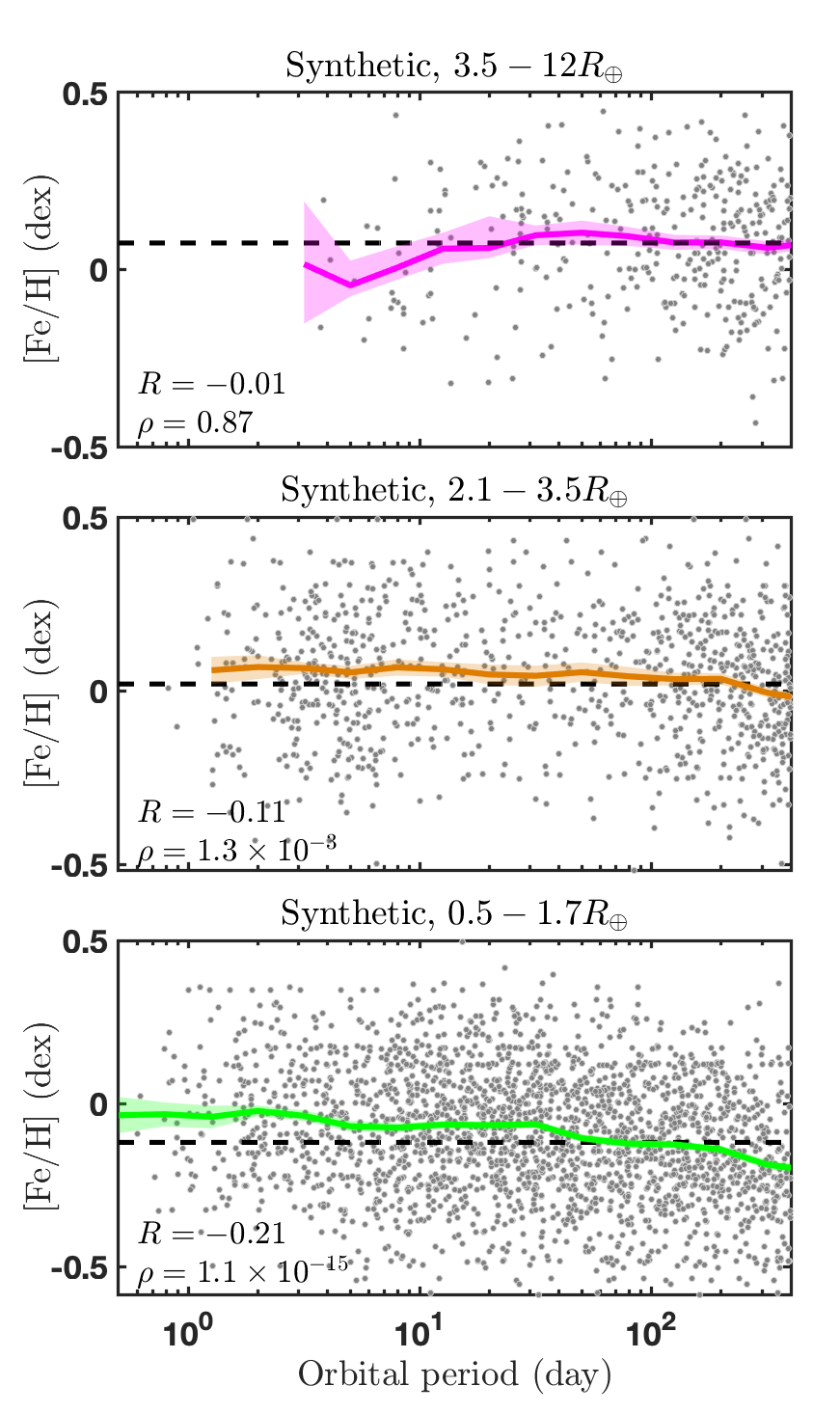}
\caption{Similar to Fig. \ref{figPeriodFeHALL} for large planets (Top), sub-Neptunes (Middle), and terrestrial planets (Bottom).
\label{figPeriodFeHdifferentradii}}
\end{figure}

\begin{figure}[!t]
\centering
\includegraphics[width=0.95\linewidth]{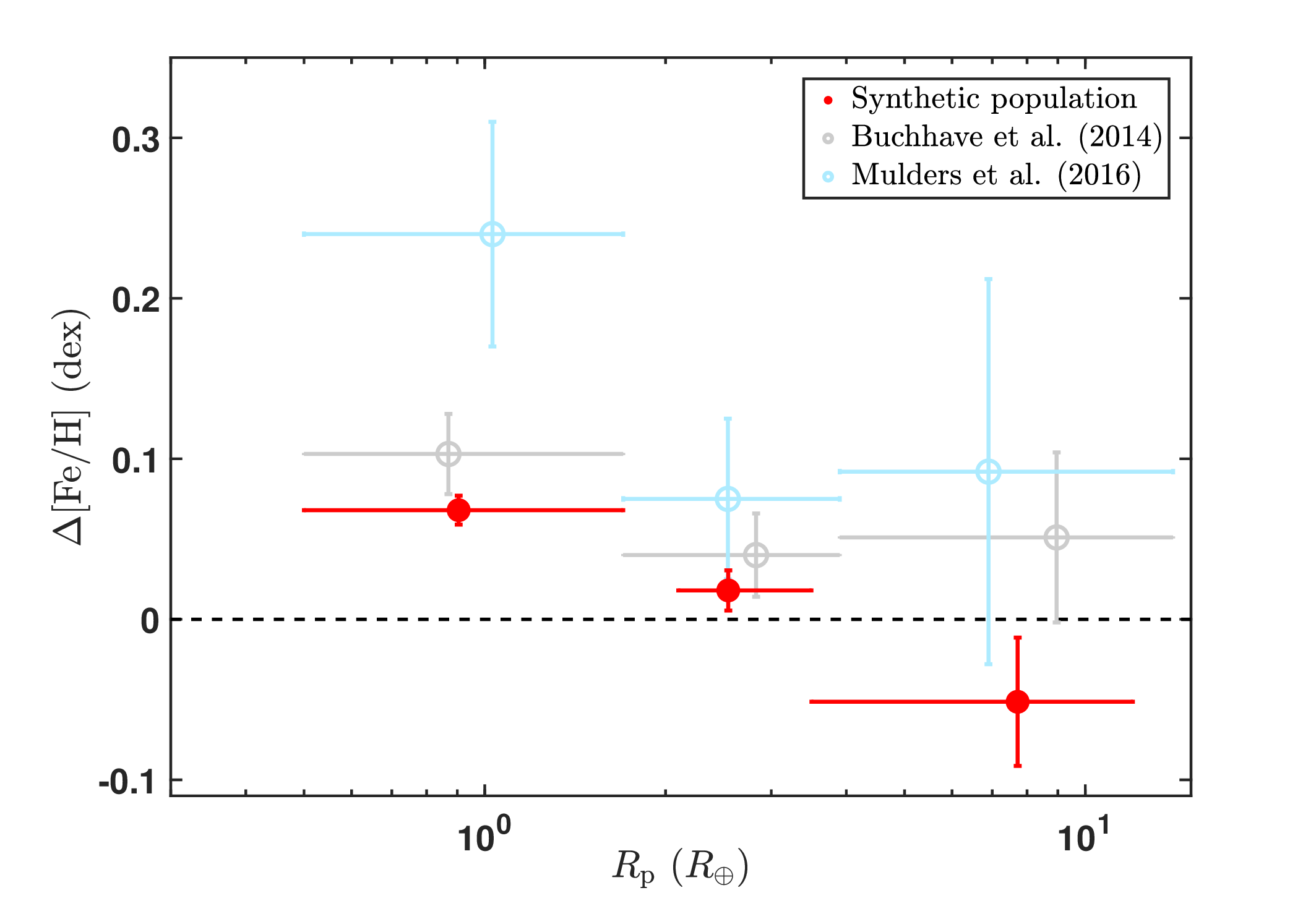}
\caption{Difference in metallicity between stars hosting planets with orbital period $P<10$ and $>10$ days.
The solid red points and error-bars show the results derived from the synthetic sample generated by the Bern model.
The grey and cyan circles denote the results from \cite{2014Natur.509..593B} and \cite{2016AJ....152..187M}, respectively.
\label{figDeltaFeHdifferentradii}}
\end{figure}

We further divide planets into three categories according to their sizes: large planets ($3.5-12 R_\oplus$), sub-Neptunes ($2.1-3.5 R_\oplus$) and terrestrial planets ($0.5-1.7 R_\oplus$). 
In Fig. \ref{figPeriodFeHdifferentradii}, we show the metallicity-period distributions for the three categories of planets.
We perform the Pearson tests and the resulted Pearson coefficients (corresponding $p-$values) are -0.01 (0.87), -0.11 ($1.3 \times 10^{-8}$) and -0.21 ($1.1 \times 10^{-15}$) for the large planets, sub-Neptunes and terrestrial planets, respectively.
We also calculate the difference in host stellar metallicity of the planets with orbital period $<10$ and $\ge 10$ days.
As shown in Fig. \ref{figDeltaFeHdifferentradii}, $\Delta \rm [Fe/H]$ derived from the synthetic sample are largest ($0.07 \pm 0.01$ dex) for terrestrial planets, smaller for sub-Neptunes ($0.02 \pm 0.01$
dex), and statistically indistinguishable from no difference for large planets
($-0.05 \pm 0.04$ dex).

\begin{figure}[!t]
\centering
\includegraphics[width=0.85\linewidth]{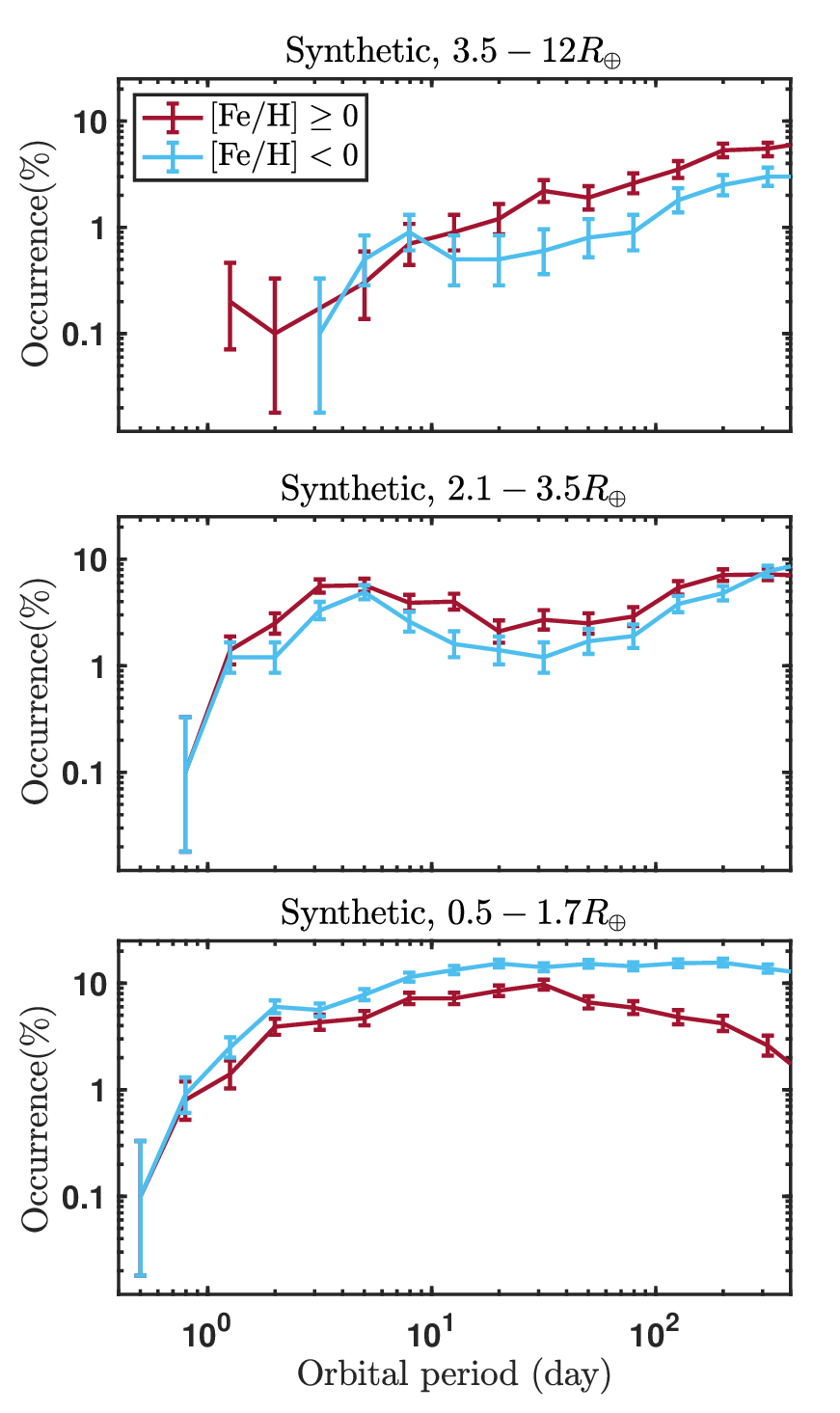}
\caption{Planetary occurrence rate as a function of orbital period for super-solar and
sub-solar metallicity stars derived from the synthetic population.
\label{figOccurrenceFeHdifferentradii}}
\end{figure}

To further investigate how the metallicity-period dependence affects the overall planet population, we divide the synthetic sample into two subsamples of super-solar ($\rm [Fe/H] \ge 0$) and sub-solar
($\rm [Fe/H]<0$) metallicities. 
Then we calculate the occurrence rates of planets of different sizes at different orbital periods for the two subsamples.
As shown in Fig. \ref{figOccurrenceFeHdifferentradii}, the occurrence rates of large planets and sub-Neptunes are generally higher for the super-solar metallicity stars, while the terrestrial planets are preferentially hosted by sub-solar metallicity stars.
To evaluate the effect of metallicity on the orbital period distribution, for the two subsamples, we calculate the relative ratio of the occurrence
rate interior and exterior to a 10 day orbital period, $\frac{F(P <10 {\rm days})}{F(P \ge 10 {\rm days}})$.
To obtain the uncertainties, we assume that the numbers of planets obey the Poisson distribution. Then we resample from the given distribution and calculate the ratios for 10,000 times. The uncertainty (1-$\sigma$ interval) of each parameter is set as the range of $50 \pm 34.1$ percentiles of the 10,000 calculations. 
For the terrestrial planets, $\frac{F(P <10 {\rm days})}{F(P \ge 10 {\rm days}})$ for the super-solar metallicity subsample ($0.45 \pm 0.04$) is significantly ($\gtrsim 3-\sigma$) higher that of sub-solar metallicity subsample ($0.29 \pm 0.02$).
In contrast, for large planets and sub-Neptunes, $\frac{F(P <10 {\rm days})}{F(P \ge 10 {\rm days}})$ for super-solar and sub-solar metallicity subsamples are statistically consistent within 1-$\sigma$ interval.

Form observations, previous studies \citep[e.g.][]{2014Natur.509..593B,2016AJ....152..187M} shows similar results, i.e. exoplanets (especially for rocky planets with $R_{\rm p}<1.7 R_\oplus$) with orbital periods less than 10 days are preferentially to be hosted by metal-rich stars (see grey and cyan circles in Fig. \ref{figDeltaFeHdifferentradii}).

One natural explanation for the anti-correlations between the orbital period and metallicity for the synthetic sample is that planets need to reach a certain mass in order to migrate inward significantly, which is easier at higher metallicities because the protoplanetary disks have more solids.
The minimum masses for migration are given by either the equality masses or the isolation masses. 

Nevertheless, quantitatively, $\Delta \rm [Fe/H]$ derived from the whole synthetic sample ($0.05 \pm 0.01$ dex) is smaller than that from the observational sample ($0.15 \pm$ 0.05 dex).
Such a discrepancy can also be seen in all the three planet categories (i.e. large planets, sub-Neptunes, and terrestrial planets).
Some potential reasons for the above discrepancies are listed as follows: 
\begin{enumerate}
    \item In the Bern III models, we do not consider the dependence of the initial inner edges of the protoplanetary disks on the metallicities.
    Previous studies show that the inner disk edge around metal-rich stars may be closer to their host stars since a larger dust-to-gas ratio could increase the disk opacity, and the dust sublimation would move inward \citep{2009A&A...506.1199K,2016ApJ...832...34M}.
    If this is the case, the position where planets form or halt their migration would be closer in around metal-rich stars. 

    \item The high-eccentricity migration pathway to form close-in planets is not included in the Bern Gen III model, as there is no tidal circularization of planetary orbits by the host stars \citep{2007ApJ...669.1298F}.

    \item The Bern model only considers interactions in the first 100 Myrs.
    Long-term planetary interactions (e.g. secular chaos) in multiple systems (favour the metal-richer stars) could deliver some additional planets (hot Jupiters, ultra-short period planets) to close orbits.
\end{enumerate}
These three effects that are not considered can further strengthen the anti-correlation between planetary period and stellar metallicity, and increase the difference between the metallicities of planets with orbital periods interior and exterior to 10 days.




\section{Orbital eccentricity vs. metallicity}
\label{sec.eccen}
Orbital eccentricity is a fundamental planetary property, which contains important clues on the evolution history of planetary architectures.
Plenty of statistical studies have found a positive correlation between stellar metallicity and planetary eccentricity both for giant planets \citep{2013ApJ...767L..24D,2018ApJ...856...37B} and smaller planets \citep{2019AJ....157..198M}.
In this section, using population synthesis, we investigate the correlation between stellar metallicity and planetary eccentricity and then compare with the observational results.

\subsection{Giant planets}
\label{sec.eccen.GP}
\cite{2013ApJ...767L..24D} and \cite{2018ApJ...856...37B} collect data for giant planets from several radial velocity (RV) surveys \citep[mainly HARPS, HIRES/KICK, FIES, and TRES][]{2008psa..conf..287F,2014AN....335...41T,2017AJ....153..208B,2020A&A...636A..74T}.
Since most of giant planets detected by RV only have measurements of minimum masses ($M_{\rm p} \sin i$), we also compute the minimum masses of planets (i.e. $\sin i$) in the synthetic population using the following procedures:
First, for each planetary system, we assume the longitude $\phi$
and latitude $\theta$ of the line of sight of observer with respect to the reference frame of the given system obey a random distribution and obtain the unit vector of the direction of observer $\hat{o}$.
Second, for a planets in a given system, we compute its orbital angular momentum with respect to the reference frame $\hat{h}$ from its inclination $i_{\rm p}$ and longitude of the ascending node $\Omega_{\rm p}$.
The formula of $\hat{o}$ and $\hat{h}$ are as follows:
\begin{equation}
    \hat{o} = \left( \begin{array}{c}
     \cos \phi \sin \theta \\
     \sin \phi \sin \theta \\
     \cos \theta
\end{array} \right) 
\mbox{~and~} \hat{h} =\left( \begin{array}{c}
     \sin \Omega_{\rm p} \sin i_{\rm p} \\
     -\cos \Omega_{\rm p} \sin i_{\rm p} \\
     \cos i_{\rm p}.
\end{array} \right)
\end{equation}
Finally, the inclination of the orbit of the planet with respect to the observer $i$ is the angle between $\hat{o}$ and $\hat{h}$. Thus,
\begin{equation}
    \cos i = \hat{o} * \hat{h}, \sin i = \sqrt{1-{\cos}^2 i}.
\end{equation}

\begin{figure}[!t]
\centering
\includegraphics[width=\linewidth]{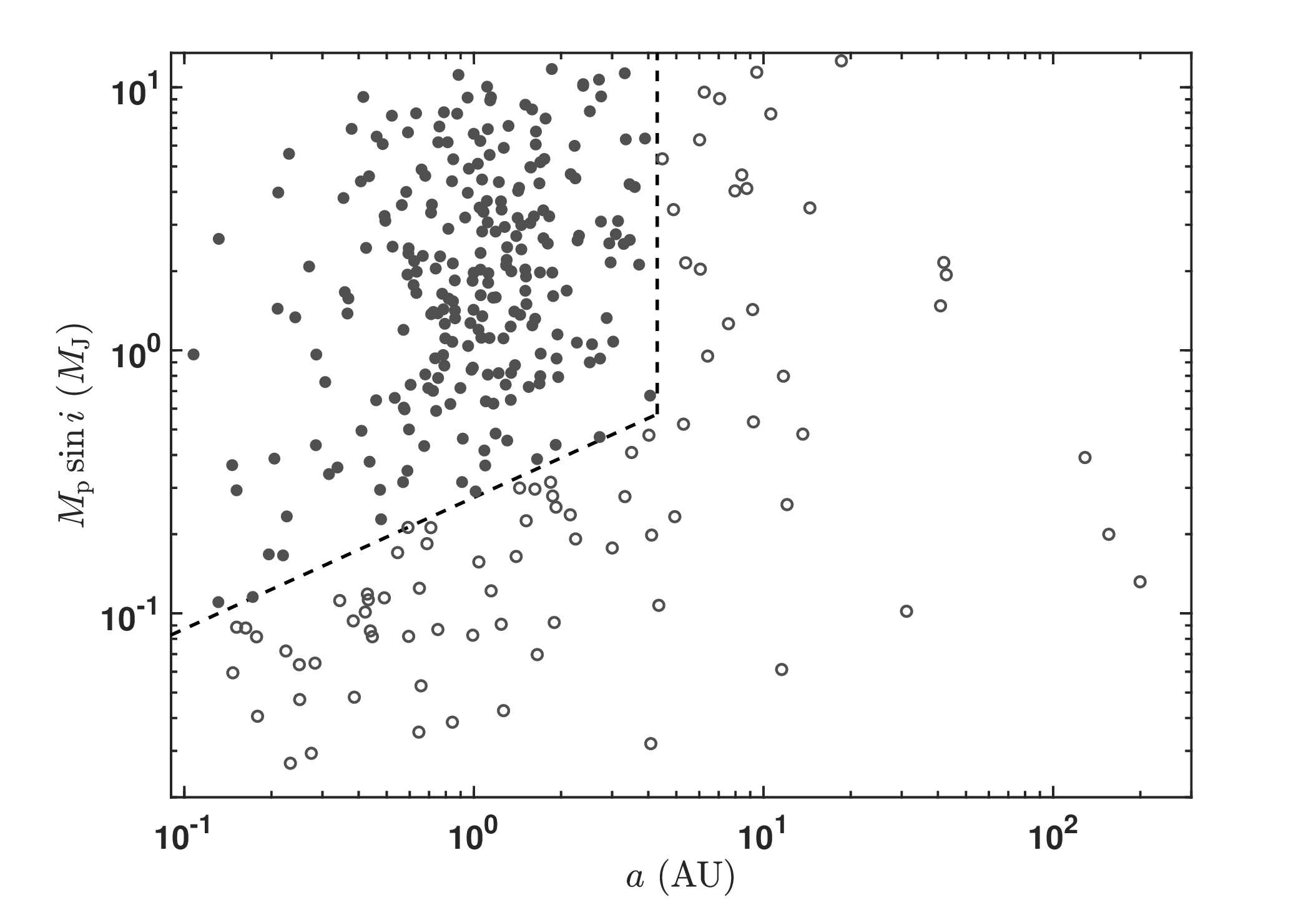}
\caption{The semi-major axis ($a$) vs. effective mass ($M_{\rm p} \sin i$) for giant planets in the biased RV synthetic population at 5 Gyr.
Solid points and open circles denote planets above and below detection limit, respectively.
The dashed line represents the typical detection limit for corresponding RV surveys when $e=0$.
\label{figGiantplanetaMp}}
\end{figure}

\begin{figure}[!t]
\centering
\includegraphics[width=\linewidth]{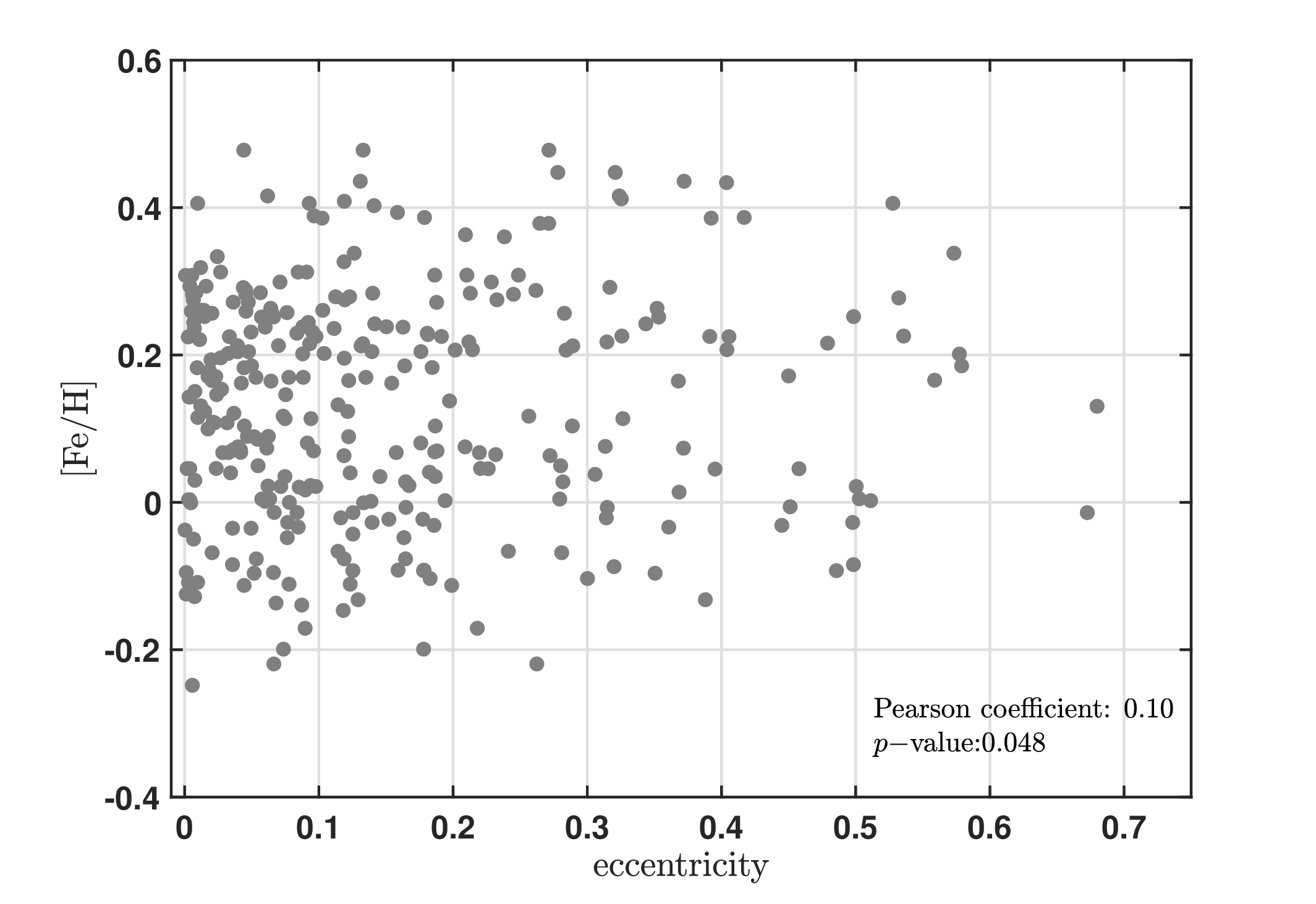}
\caption{The planetary orbital eccentricity ($e$) vs. stellar metallicity ($\rm [Fe/H]$) for giant planets in the synthetic population at 5 Gyr after selection according to the RV detection bias.
We also print the Pearson coefficient and corresponding $p-$value.
\label{figGiantplaneteFeH}}
\end{figure}

With similar criteria as these works, we select planets with minimum masses in the range of $0.1-13$ Jupiter mass ($M_{\rm J}$) as giant planets from the synthetic population at 5 Gyr.
To avoid the influence of tides \citep{2024AJ....168..115B}, we only considered planets with orbital semi-major axes $a$ larger than $0.1$ AU \citep[e.g.][]{2008ApJ...678.1396J,2012MNRAS.423..486L}.
Furthermore, to apply the similar detection bias as previous statistical studies \citep{2013ApJ...767L..24D, 2018ApJ...856...37B}, we download data from RV surveys \citep[mainly HARPS, HIRES/KICK, FIES, and TRES;][]{2008psa..conf..287F,2014AN....335...41T,2017AJ....153..208B,2020A&A...636A..74T} and obtain the median values of the root mean squares (RMS) of the semi-amplitudes and time baselines for Sun-like (i.e. FGK-type main-sequence) stars.
Then we only kept giant planets with $K> {\rm RMS_{median}} \ \& \ P<\rm time \ baseline$, where $P$ is the orbital period and $K = \frac{28.4 {\rm m \ s^{-1}}}{\sqrt{1-e^2}} \times \frac{M_{\rm p} \sin{i}}{M_{J}} \left(\frac{P}{1 \rm yr}\right)^{-1/3} \left(\frac{M_*}{M_\odot}\right)^{-2/3}$ is the amplitude of the generated RV signal.
After applying the above selections, we are left with 155 stars hosting 240 planets.
Figure \ref{figGiantplanetaMp} displays the $M_{\rm p} \sin i-a$ diagram of giant planets in our sample.


\begin{figure}[!t]
\centering
\includegraphics[width=\linewidth]{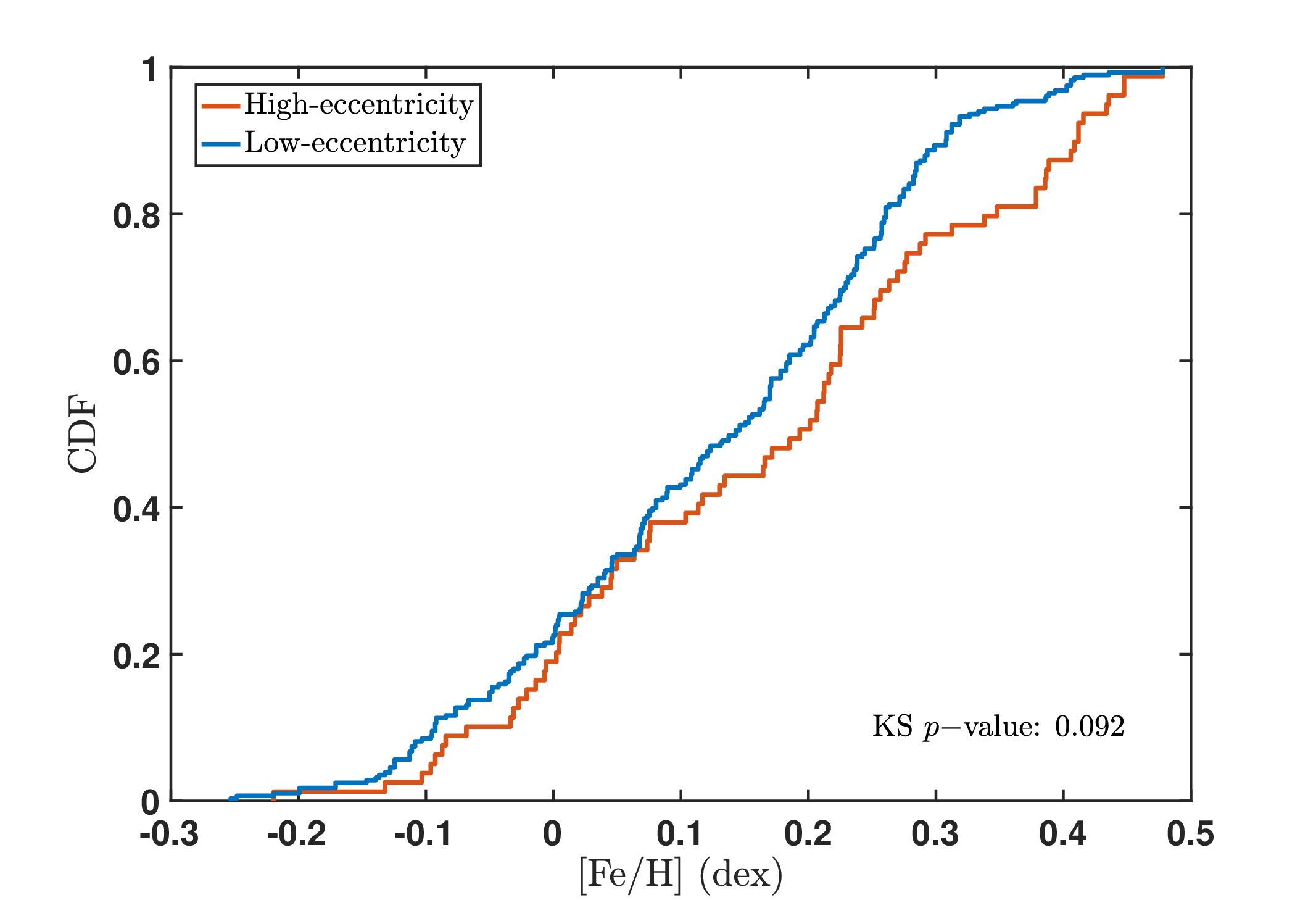}
\caption{The cumulative distributions of stellar metallicities $\rm [Fe/H]$ for high-eccentricity (red) and low-eccentricity giant planets (blue).
The two sample $KS$ test $p-$value is printed at the corner.
\label{figCDFFeHGPcategories}}
\end{figure}

Figure \ref{figGiantplaneteFeH} shows the planetary eccentricity-stellar $\rm [Fe/H]$ diagram for giant planets in the biased RV sample (above the detection limit).
We perform the Pearson test and the resulting to a Pearson coefficient of 0.10 with a $p-$ value of 0.048, suggesting a positive eccentricity-metallicity correlation for giant planets with $\sim 2-\sigma$ confidence.
Referring to previous studies \citep{2013ApJ...767L..24D, 2018ApJ...856...37B}, we further divide them into two categories according to their eccentricities:
\begin{enumerate}
\item high-eccentricity $e>0.25$ (46);
\item low-eccentricity $e \le 0.25$ (194);
\end{enumerate}
As shown in Fig. \ref{figCDFFeHGPcategories}, the stars hosting high-eccentricity giant planets have a larger distribution in $\rm [Fe/H]$ comparing to those hosting low-eccentricity giant planets.
To evaluate the significance, we perform the two-sample KS tests.
The resulting $p-$values are 0.092,
suggesting high-eccentricity giant planets are preferentially hosted by metal-richer stars compared to low-eccentricity giant planets with confidence level of $\sim 1-2\sigma$.
The average metallicities for stars hosting high-eccentricity and low-eccentricity giant planets are $0.17 \pm 0.02$ and $0.12 \pm 0.02$ dex (uncertainties derived from bootstrapping), respectively. 

\begin{figure}[!t]
\centering
\includegraphics[width=\linewidth]{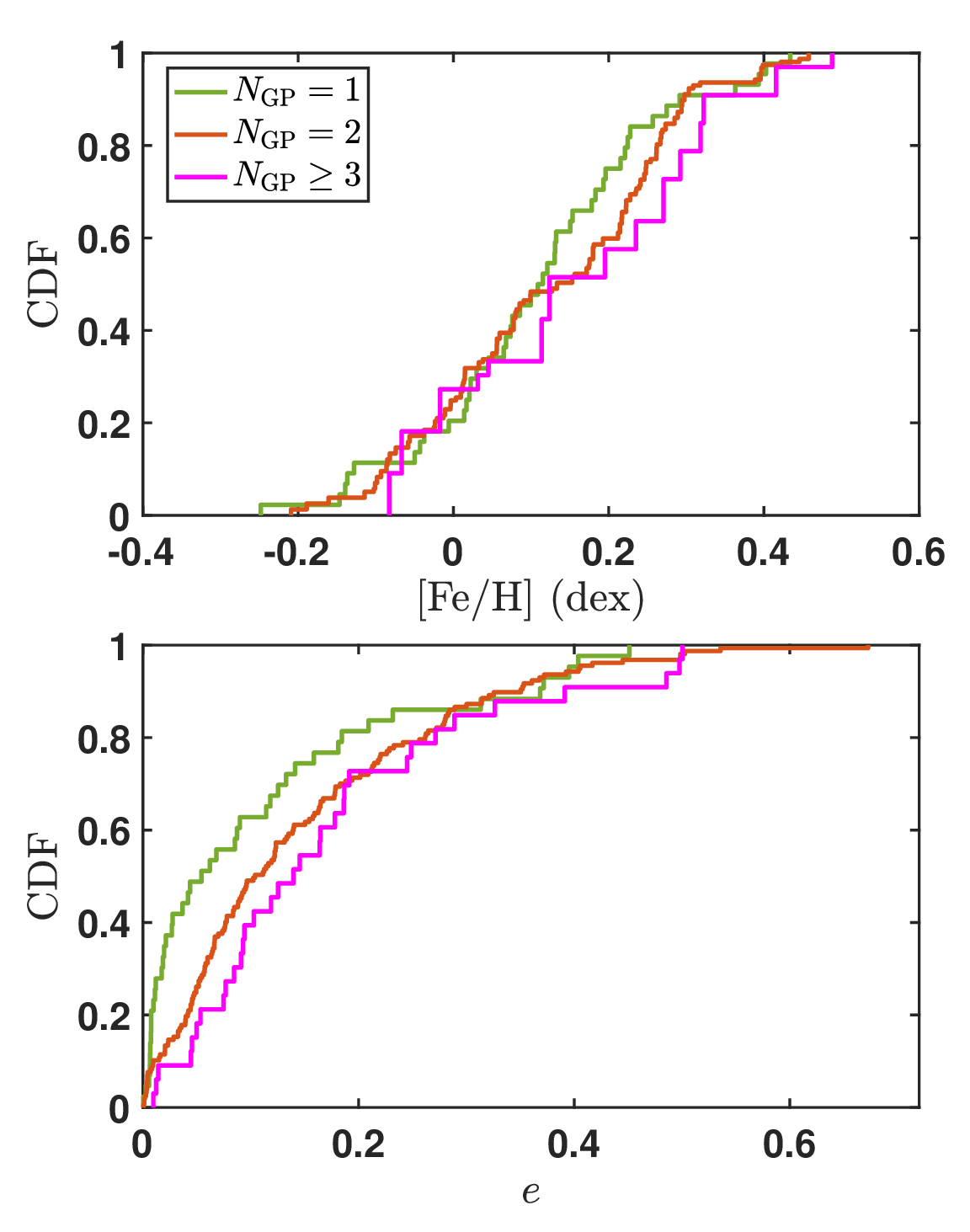}
\caption{The cumulative distributions of stellar metallicities $\rm [Fe/H]$ (Top panel) and planetary eccentricity (Bottom panel) for single giant planetary systems (blue) and multiple giant planetary systems.
The two sample KS test $p-$values are printed at the lower-right corner of each panel.
\label{figGantplaneteFeH_singlevsmultiple}}
\end{figure}

Previous studies have proposed a potential explanation for the eccentricity-metallicity trends:
metal-richer stars are more likely to form more giant planets and perturbations of outer giant planets could have pump up planetary eccentricities \citep{2018ApJ...856...37B}.
To examine this explanation, we compare the distributions in $\rm [Fe/H]$ for stars hosting 1, 2, and 3+ giant planets (regardless of whether giant planets could be detected or not). 
As shown in the top panel of Fig. \ref{figGantplaneteFeH_singlevsmultiple}, stars hosting more giant planets have relatively higher $\rm [Fe/H]$.
To evaluate the significance, we
performed the two-sample KS tests and the resulted $p-$values are 0.185 and 0.129 for stars hosting 2 and 3+ giant planets compared to those with 1 giant planets,
corresponding to confidence levels of $\sim 1-\sigma$.
In the bottom panel of Fig. \ref{figGantplaneteFeH_singlevsmultiple}, we then compare the distribution of the eccentricities of synthetic giant planets in systems with 1, 2 and 3+ giant planets.
As can be seen, single giant planets have a smaller eccentricity distribution compared to those in systems with 2 and 3+ giant planets with KS $p-$values of 0.1855 and 0.0732.
The higher stellar $\rm [Fe/H]$ and larger planetary eccentricity (though statistically insignificant) of giant planets accompanied by other giant planets are generally in agreement with the expectation and thus support that the proposed mechanism could help to explain why eccentric giant planets favour metal-rich stars.

The above synthetic results suggest that giant planets with high eccentricity tend to be hosted by metal-richer stars, which is generally consistent with the statistical results \citep[see Fig. 2 of][]{2018ApJ...856...37B}.
Nevertheless, compared to the synthetic sample, the observed high-eccentricity (low-eccentricity) giant planets have significantly larger (smaller) metallicities, with an average of $0.23 \pm 0.04$ ($-0.07 \pm 0.05$), leading to a more significant correlation between eccentricity and metallicity in the observations.
Such a discrepancy is not unexpected.
Due to computational cost constraints, we only simulate the interactions between planets for the first 100 Myr.
Long-term perturbations between giant planets can further excite their eccentricities.
This could lead to a stronger eccentricity-metallicity correlation since giant planets tend to be hosted by metal-richer stars.
Moreover, as discussed in Sect. 5.5 of \citet{2023EPJP..138..181E}, the final number of giant planets in a given system  may not be  same as the total number originally formed, due to collisions and ejections (see their Fig. 16).


\subsection{Small planets}
\label{sec.eccen.SP}
\begin{figure}[!t]
\centering
\includegraphics[width=\linewidth]{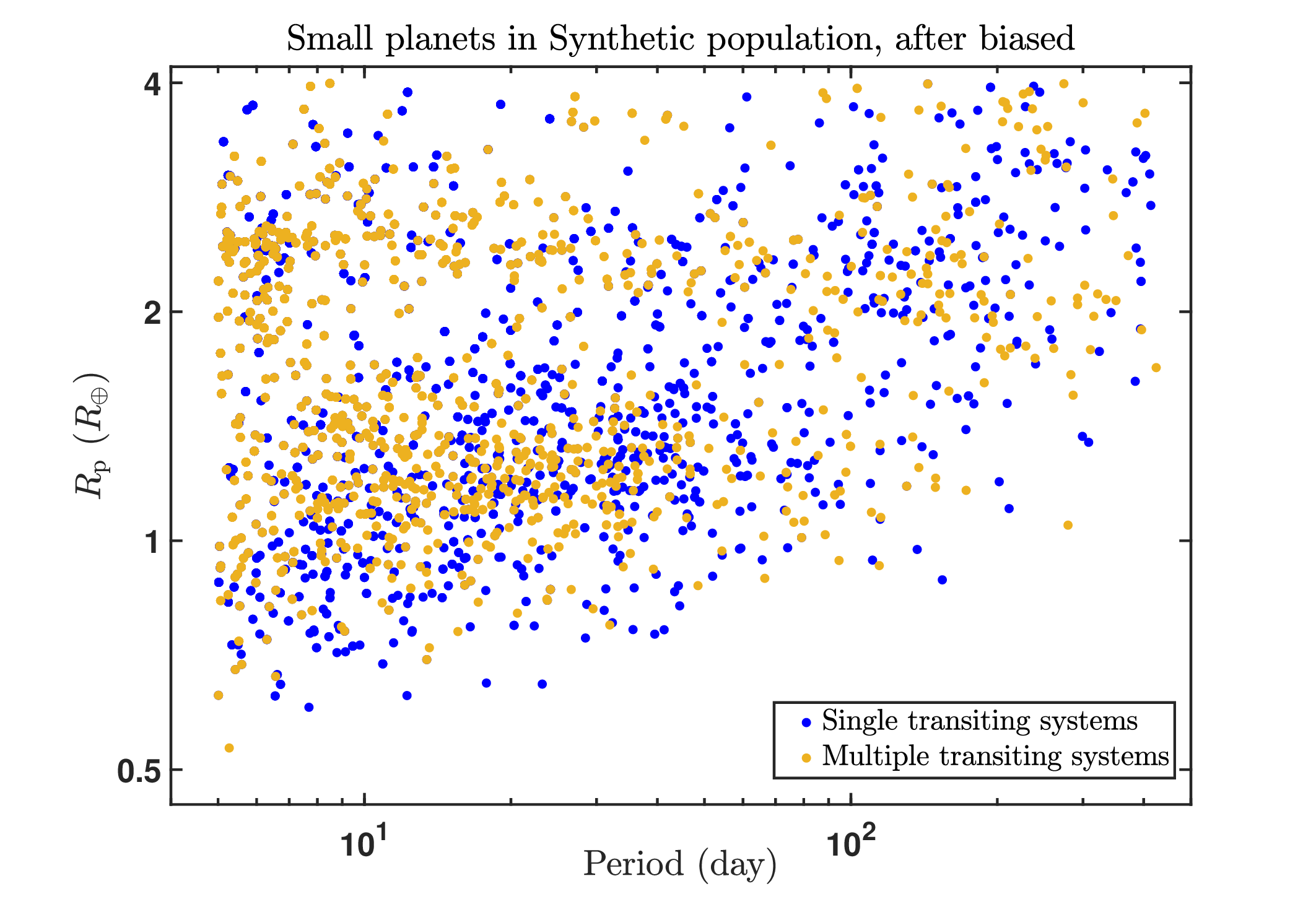}
\caption{The orbital period ($P$) vs. radius ($R_{\rm p}$) for selected small planets in the Kepler-biased synthetic population at 5 Gyr.
\label{figSmallplanetsPRp}}
\end{figure}

For the small planets with radii less than $\sim 4 R_\oplus$, \cite{2019AJ....157..198M} provides tentative evidences that eccentric small planets with radii between $1.4-4 R_\oplus$ favor metal-rich stars using the CKS sample.
Recently, \cite{2023AJ....165..125A} confirmed
a significant eccentricity-metallicity trend for Kepler planets/candidates with $R_{\rm p} <4 R_\oplus$ using LAMOST-Gaia-Kepler sample.
In order to compare synthetic results with Kepler observational results, we first select planets with $R_{\rm p}<4 R_\oplus$ and orbital period $P>5$ days with the same criteria as \cite{2023AJ....165..125A} and then apply the detection biases using KOBE program \citep{2021A&A...656A..74M}.
Figure \ref{figSmallplanetsPRp} shows the period-radius diagram for selected small planets in the Kepler-biased synthetic population.

\begin{figure}[!t]
\centering
\includegraphics[width=\linewidth]{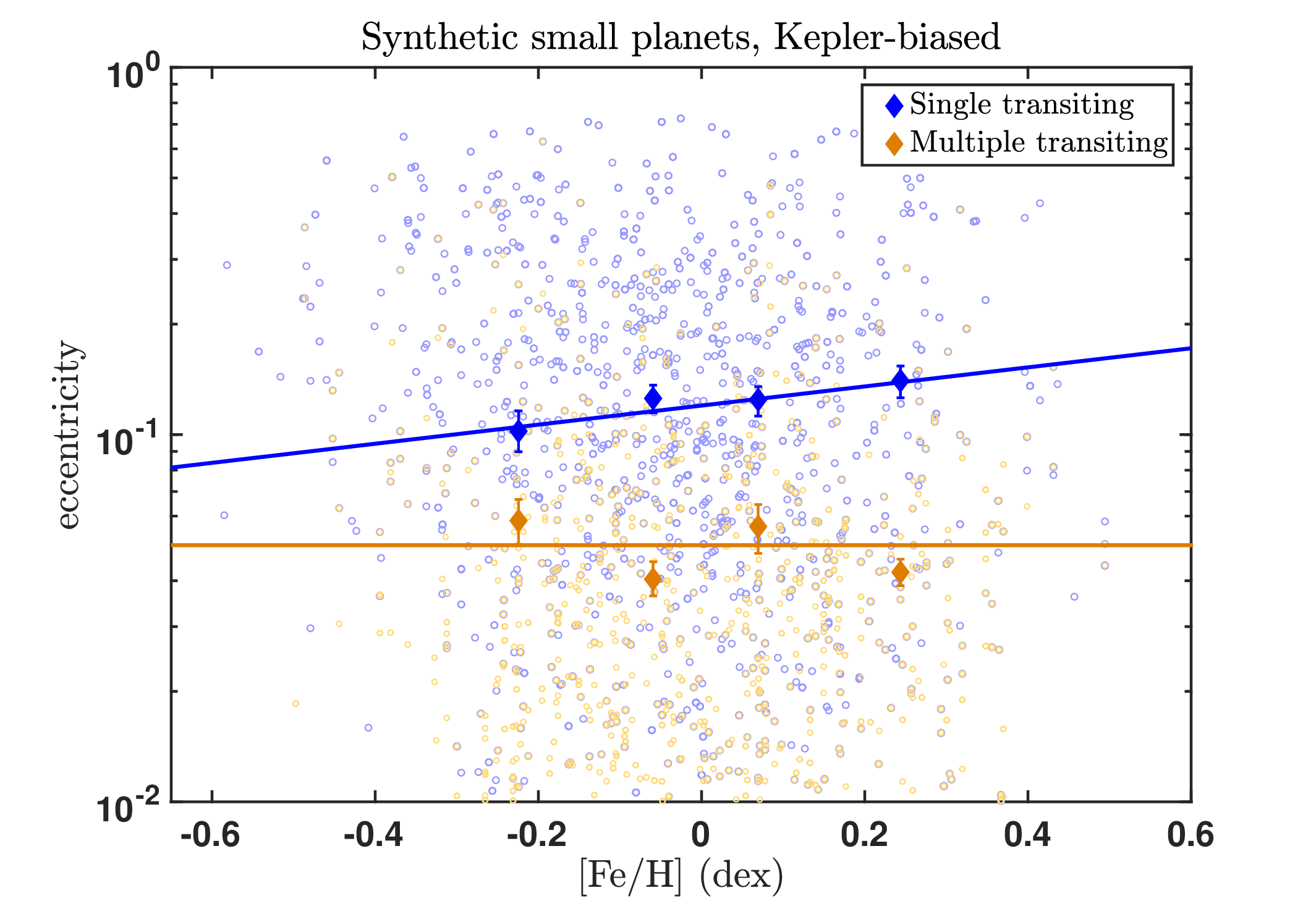}
\caption{The orbital eccentricity vs. stellar metallicity ($\rm [Fe/H]$) for Kepler-detectable small planets in single (blue) and multiple (orange) systems in the synthetic population.
We also plot the mean planetary eccentricity and their 1-$\sigma$ interval ($50 \pm 34.1\%$ percentiles) from bootstrapping as solid diamonds and errorbars.
The best fits are plotted as solid lines.
\label{figSmallplanetseFeH}}
\end{figure}

Based on the selected synthetic sample, we investigate the orbital eccentricity $e$ of small planets in single and multiple systems as a function of stellar $\rm [Fe/H]$.
Figure \ref{figSmallplanetseFeH} displays the stellar $\rm [Fe/H]$-planetary eccentricity diagram for small planets in single (blue) and multiple (orange) systems.
As shown in Fig. \ref{figSmallplanetseFeH}, small planets in multiple systems are mostly on nearly circular orbits (with $e \lesssim 0.1$), while small planets in single systems have a larger eccentricity distributions.

We adopt the two-sample KS test for the Kepler-biased synthetic sample and find that the eccentricity of small planets in single systems are significantly larger than those in multiple systems with a $p-$value of $7.9 \times 10^{-130}$, which is consistent with the `eccentricity-dichotomy' revealed by Kepler data \citep{2015ApJ...808..126V,2016PNAS..11311431X,2019AJ....157..198M}.
Quantitatively, the average eccentricities of single systems ($\bar{e} \sim 0.140$) and multiple systems ($\bar{e} \sim 0.05$) from the synthetic sample are both well consistent with the observational results from California-Kepler Survey data, i.e. Rayleigh distributions with $\sigma_e$ of $0.167^{+0.0013}_{-0.0008}$ and $0.035 \pm 0.006$ \cite{2019AJ....157..198M}.  
\cite{2016PNAS..11311431X} derive a slightly large average eccentricity ($\sim 0.3$) for planets in single systems and \cite{2019AJ....157..198M} suggest that this discrepancy could be that the stellar properties using in \cite{2016PNAS..11311431X} have not be improved by the Gaia catalogues and result in a slightly higher eccentricity distribution.

Moreover, with the increase of stellar $\rm [Fe/H]$,
the eccentricities of small planets in single system grow, while the eccentricity of small planets in multiple systems changes mildly.
To quantify the correlation between $e$ and $\rm [Fe/H]$, following \cite{2023AJ....165..125A}, we then divide the small planets in the synthetic sample into four bins according to their $\rm [Fe/H]$ and calculate their average orbital eccentricities $\bar{e}$, which is shown in Fig. \ref{figSmallplanetseFeH}. 
Then we fit $\bar{e}$ and $\rm [Fe/H]$  with a constant model ($\bar{e} = \rm constant$) and an exponential model ($\bar{e} = a \times 10^{(b \times \rm [Fe/H]})$) with the Levenberg-Marquardt algorithm (LMA).
For the small planets in single systems, the exponential model is preferential compare to constant model with an AIC difference ($\rm \Delta AIC$) of 6.5.
In contrast, for planets in multiple systems, the constant model is preferred with a smaller AIC score ($\rm \Delta AIC = -0.8$).
In order to obtain the uncertainties of $\bar{e}$ and evaluate the significance,
we generate the 1,000 bootstrapped sample from the original sample and calculate the mean eccentricities for the four bins.
The uncertainty is set as the $50 \pm 34.1$ percentiles in the resampled distribution.
Then we refit the $\bar{e}- \rm [Fe/H]$ relation with the same procedure before. Out of the 1,000 sets of resampled data, the exponential model is preferred with a smaller AIC score for 961 times for the small planets in single systems (corresponding to a confidence level of 96.1\%), while for those in multiple systems, the constant model is preferred for 816 times.
The fitting results of the preferred models are:
\begin{eqnarray}
\bar{e} = \begin{cases} {0.12^{+0.01}_{-0.01} \times 10^{(0.26^{+0.08}_{-0.07} \times \rm [Fe/H])}}&\mbox{ Single}\\ 
{0.05^{+0.01}_{-0.01}}&\mbox{Multiple}.
\end{cases}
\end{eqnarray}

Referring to \cite{2019AJ....157..198M}, we further divide small planets into two subsamples, i.e. radii larger/less than 1.4 $R_\oplus$. 
We find that $e$ of planets in single systems exists positive correlations with $\rm [Fe/H]$ with Pearson coefficients of 0.11 (corresponding $p-$value of $1.1 \times 10^{-8}$) and 0.03 (corresponding $P-$value of 0.010) for the two subsamples of $R_{\rm p} \ge$ and $<1.4 R_\oplus$, respectively.
The exponential model is preferential compare to constant model with AIC difference of 4.6 and 6.3 for the two subsamples of $R_{\rm p} \ge 1.4 R_\oplus$ and $1.4 R_\oplus$ respectively, corresponding to confidence levels of $\gtrsim 2 \sigma$). The fitting results of the exponential models are:
\begin{eqnarray}
\bar{e} = \begin{cases} {0.11 \times 10^{(0.4 \times \rm [Fe/H])}}&\mbox{$R_{\rm p} \ge 1.4 R_\oplus$, Single}\\ 
{0.22 \times 10^{(0.2 \times \rm [Fe/H])}}&\mbox{$R_{\rm p} < 1.4 R_\oplus$, Single}. \end{cases}
\end{eqnarray}. 
In contrast, for planets in multiple systems, the correlations with $\rm [Fe/H]$ are statistically indistinguishable with Pearson coefficients of only 0.01 (corresponding $p-$value of 0.63) and -0.06 (corresponding $p-$value of 0.07) for the two subsamples, respectively.
Moreover, the constant model is preferred with smaller AIC score for both the two subsamples. 

Based on the above trends, we conclude that on average, small planets in single systems are on eccentric orbits and their eccentricities increase with $\rm [Fe/H]$, while small planets in multiple systems are mostly on near-circular orbits and their eccentricities shows no (significant) dependence on $\rm [Fe/H]$.
That is to say, the `eccentricity dichotomy' becomes more significant with increasing $\rm [Fe/H]$.

\begin{figure}[!t]
\centering
\includegraphics[width=\linewidth]{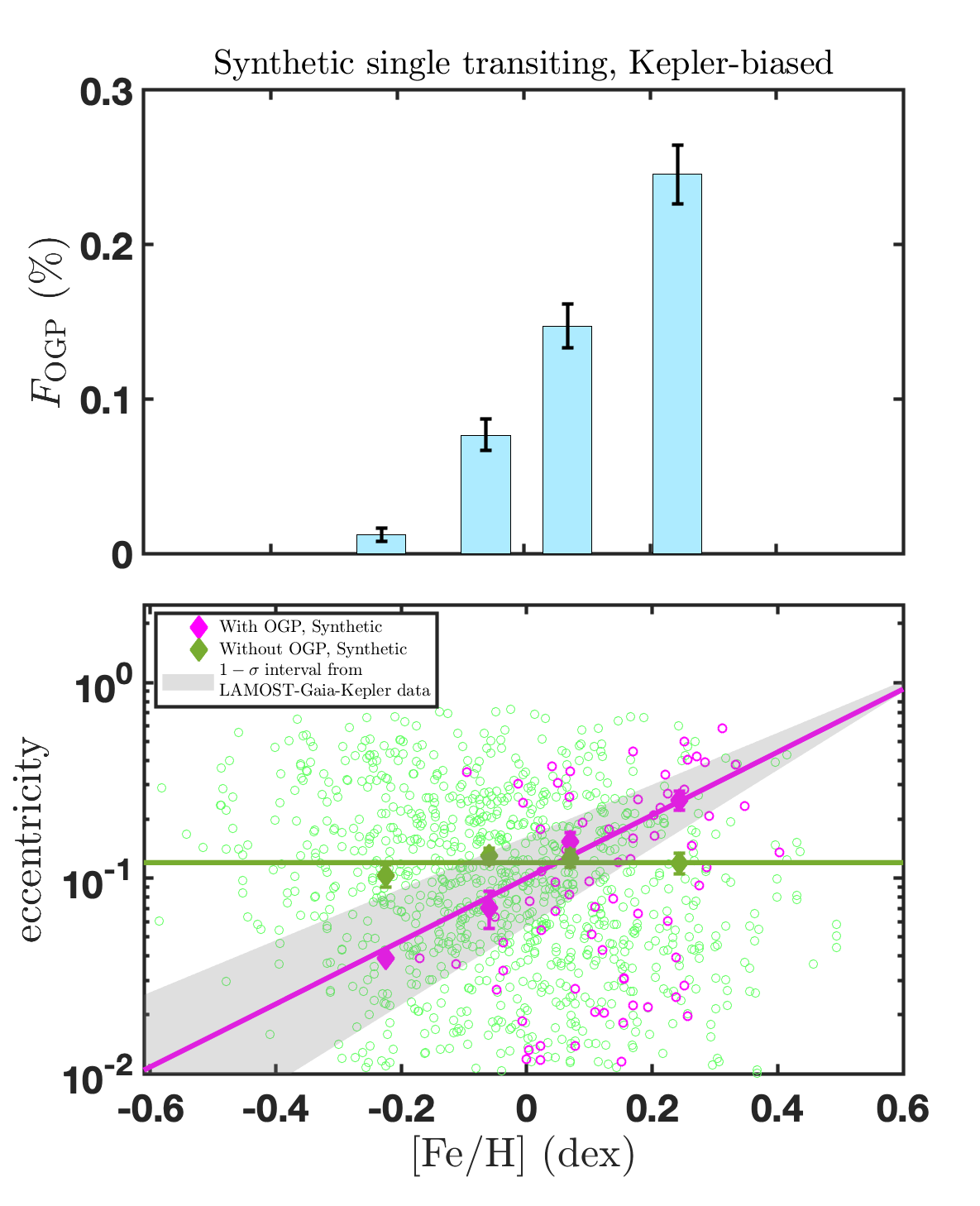}
\caption{Top: The fraction of single transiting small planets with outer giant planets (OGP) vs. stellar metallicities derived from the Kepler-biased synthetic sample.
Bottom: The orbital eccentricity vs. stellar metallicity for small transiting planets with/without OGPs (purple/green).
The solid lines represent the corresponding best fits.
The grey region denotes the $1-\sigma$ interval of the eccentricity-$\rm [Fe/H]$ correlation derived from the LAMOST-Gaia-Kepler sample \citep{2023AJ....165..125A}.
\label{figSmallplanetssingle_eccFeH}}
\end{figure}

One potential explanation for the positive eccentricity-metallicity relation of small planets in the single transiting systems is that protoplanetary disks with higher metallicities have more solid materials to form more giant planets, and the gravitational perturbations by outer giant planets would pump up larger orbital eccentricities of inner small planets \citep{2017AJ....153..210H,2018MNRAS.478..197P,2020MNRAS.498.5166P}.
To examine the above explanation, we classify these single transiting small planets into two categories: with and without outer giant planets (OGP).
In the top panel of Fig. \ref{figSmallplanetssingle_eccFeH}, we show the fraction of single transiting planets  with OGPs $F_{\rm OGP}$ as a function of stellar metallicity. 
As can be seen, $F_{\rm OGP}$ grows from $1\%$ to $30\%$ when $\rm [Fe/H]$ increases from $~-0.2$ to $0.2$, which is expected since giant planets are more preferred to be hosted by metal-richer stars compared to small planets (see Fig. \ref{figFHJCJWJ_FeH_Synthesis} and \ref{figFKep_FeH_Synthesis}).
Then we divide these two categories into four bins according to $\rm [Fe/H]$ with the same interval shown in Fig. \ref{figSmallplanetseFeH} and calculate the mean eccentricities and corresponding uncertainties.
As can be seen in the bottom panel of Fig. \ref{figSmallplanetssingle_eccFeH}, there exists an obviously positive correlation between $e- \rm [Fe/H]$ for single transiting planets with OGPs. 
In contrast, the eccentricities of single transiting planets without OGPs are more dispersed and show no significant dependence on metallicity.

To qualify the $\bar{e}- \rm [Fe/H]$ correlation for single transiting planets with/without OGPs, we fit them by adopting the same procedure as before.
For single transiting planets with OGPs, the exponential model is preferred with $\rm \Delta AIC$ of 13.1, corresponding to a confidence level of 99.9\% from the bootstrapping.
In contrast, the constant model is preferred with a smaller AIC score ($\rm \Delta AIC = -0.96 $) for the single transiting planets without OGPs. 
The best fits are:
\begin{eqnarray}
\bar{e} = \begin{cases} {0.10^{+0.01}_{-0.01} \times 10^{(1.61^{+0.21}_{-0.15} \times [\rm Fe/H])}}&\mbox{ Single with OGPs}\\ 
{0.12^{+0.01}_{-0.01}}&\mbox{Single without OGPs}.
\label{eqeccFeH_OGP}
\end{cases}
\end{eqnarray}
Out of the 1,000 sets of resampled data, the exponential indexes of single small planets with OGPs are all larger than those of small planets without OGPs, suggesting that the perturbations by outer giant planets help to excite the orbital eccentricities of inner small planets.
Moreover, the best fit of the $\bar{e}- \rm [Fe/H]$ relation for the single planets with OGPs in the synthetic population is well consistent with the result derived from the LAMOST-Gaia-Kepler sample \citep[$\bar{e} = 0.11^{+0.06}_{-0.06} \times 10^{(1.78^{+1.49}_{-0.89} \times \rm [Fe/H])}$;][]{2023AJ....165..125A}.
The above analyses demonstrate that the external eccentricity exciting mechanism could explain the positive eccentricity-metallicity correlation for inner small planets accompanied by OGPs.

\begin{figure}[!t]
\centering
\includegraphics[width=\linewidth]{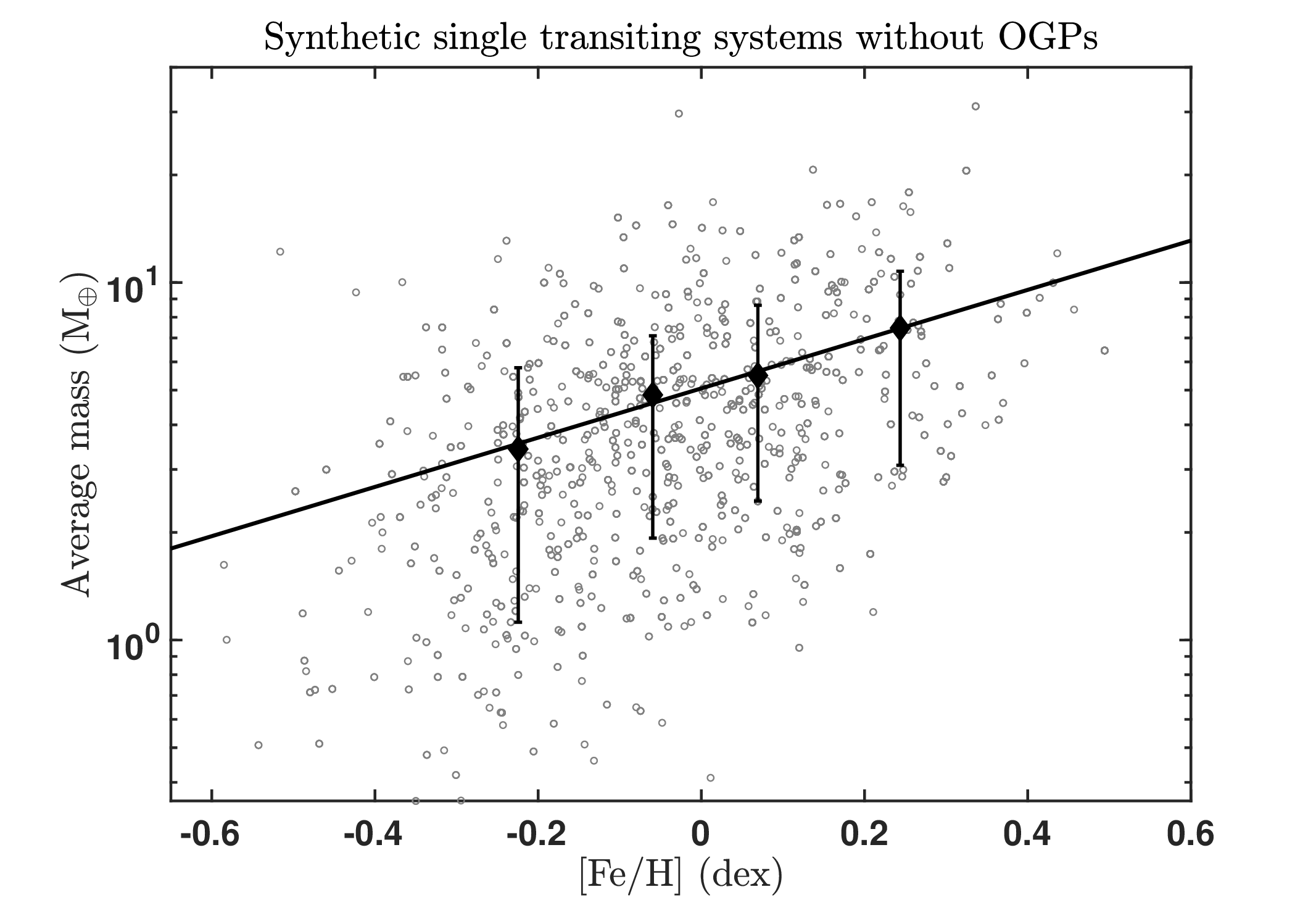}
\caption{The average masses of small planets (regardless whether could be detected or not) in single transiting systems without OGPs vs. stellar metallicity (grey circles).
The diamonds and errorbars denote the mean and 1-$\sigma$ ($50 \pm 34.1$) interval of the average masses in different bins of $\rm [Fe/H]$.
The solid line represents the best fit.
\label{figsmallplanetFeHmass}}
\end{figure}

Nevertheless, for those with no OGPs, their eccentricities are independent with stellar $\rm [Fe/H]$, which is inconsistent with the observational results.
This is not unexpected since our model only consider the interaction between planets for the first 100 Myr and thus does not include the long-term formation and evolution of small planets, which are discussed as follows:
\begin{enumerate}
    \item For single transiting small planets without OGPs in the synthetic sample, nearly all (96.9\%) are accompanied by at least one undetectable small planet ($1-4 R_\oplus$, $5<P<400$ d) according to Kepler's detection limits. In contrast, the majority (73\%) of single transiting small planets with OGPs lack any small planet companions.
    Theoretical studies and simulations have demonstrated that long-term dynamical interactions among inner small planets can pump up the orbital eccentricity/inclination of planets \citep[e.g.][]{2007ApJ...666..423Z,2016ApJ...832...34M},which is also supported by observations, as the Kepler sample reveals that planetary systems become dynamically hotter (with higher eccentricity/inclination dispersion) as stars age \citep[e.g.][]{2021AJ....162..100C,2023AJ....166..243Y}.
    The average eccentricity of small planets, induced by long-term perturbations of nearby planets, is estimated to be \citep{2007ApJ...666..423Z}:
    \begin{equation}
       <e^2>^{1/2} \approx 5.2 k_0^{-3/2} \mu^{1/2} {\left(\frac{t}{\mathrm{yr}}\right)}^{1/2},
   \end{equation}
   where $k_0$ is the scaled initial separation. $\mu$ is the average mass of small planets in one system.
   Figure \ref{figsmallplanetFeHmass} displays the average masses of small planets (regardless whether could be detected or not) in single transiting systems without OGPs as a function of metallicity.
   As can be seen, the average masses $\mu$ show an obviously positive correlation with $\rm [Fe/H]$ (Pearson coefficient of 0.39 with a $p-$ value of $1.0 \times 10^{-130}$) and 
   $\mu \sim 5 M_\oplus \times 10^{(0.7 \times \rm [Fe/H])}$, which would lead to an increase in the exponential index by $\sim 0.35$. 

   \item Some small planets may even experience merge/ejection due to the interaction of close planets \citep[e.g.][]{2015ApJ...807...44P}, especially for planets with lower masses, since they are expected to have larger amplitudes of the eccentricity \citep[e.g.][]{2011ApJ...739...31L}.
   This would remove some eccentric planets ($e>0.3$) with lower masses (mostly with $R_{\rm p} < 1.7 R_\oplus$) around stars with sub-solar metallicities ($\rm [Fe/H] \lesssim -0.1$ dex) in the synthetic sample, which would further increase the exponential index and help to explain why such planets are absent in the observed sample (see fig. 6, 8 and 10 of \cite{2023AJ....165..125A}, $e > 0.3$ are beyond 3-$\sigma$ for stars of $\rm [Fe/H] \lesssim -0.1$).
\end{enumerate}




\begin{table*}[!t]
\centering
\renewcommand\arraystretch{1.5}
\caption{Dependence of the planetary properties as a function of stellar $\rm [Fe/H]$ derived from the synthetic \texttt{NG76Longshot} population.}
{\footnotesize
\label{tab:FeHcorrelations}
\begin{tabular}{l|c|c} \hline
    & Synthetic results & Comparison with observations \\ \hline 
   \multicolumn{3}{c}{Sect. \ref{sec.frequency}: Occurrence rate vs. $\rm [Fe/H]$} \\ \hline
   Giant planets (Fig. \ref{figFHJCJWJ_FeH_Synthesis}) & Positively-correlated as $F \propto 10^{\sim 1.3 \times \rm [Fe/H]}$ & Quantitatively consistent \\ 
   Neptune-sized  (Fig. \ref{figFHNCNWN_FeH_Synthesis}) & Positively-correlated as $F \propto 10^{\sim 0.5-1 \times \rm [Fe/H]}$ & Quantitatively consistent \\
   Small planets ($1-3.5 R_\oplus$)  (Fig. \ref{figFKep_FeH_Synthesis}) & First increase before $\rm [Fe/H] \lesssim 0.1$ dex and then decrease & Qualitatively consistent \\
   Sub-Earths (Fig. \ref{figFSubEarth_FeH_Synthesis}) & Anti-correlated as $F \propto 10^{-1.9 \times \rm [Fe/H]}$ & Awaiting further observational results \\ \hline
   \multicolumn{3}{c}{Sect. \ref{sec.valley}: Radius valley morphology vs. $\rm [Fe/H]$ (Fig. \ref{figRadiusValleyFeH}, \ref{figRadiushistFeH}, \ref{figRadiusmetricFeH})} \\ \hline
   $C_{\rm Valley}$ (Eq. \ref{eqDV}) &  Positively-correlated (Eq. \ref{eqC_Fe}) & \multirow{5}{*}{All quantitatively consistent} \\
   $C_{\rm Valley}$ (Eq. \ref{eqREN}) &  Positively-correlated (Eq. \ref{eqC_Fe}) &  \\
   $R^{+}_{\rm Valley}$ (Eq. \ref{eqRNSE}) &  Positively-correlated (Eq. \ref{eqR+_Fe}) &  \\
   $R^{-}_{\rm Valley}$ (Eq. \ref{eqRSE}) &  not correlated (Eq. \ref{eqR-_Fe}) &  \\
   $f_{\rm NP}$ (Eq. \ref{eqfNep}) & Positively-correlated (Eq. \ref{eqfNp_Fe}) &  \\ \hline
   \multicolumn{3}{c}{Sect. \ref{sec.distribution_radii}: Orbital periods vs. $\rm [Fe/H]$} \\ \hline
   All planets & higher $\rm [Fe/H]$ at short period ($\rm \Delta [Fe/H] = 0.05 \pm 0.01$, Fig. \ref{figPeriodFeHALL}) &  \\
   Large planets & No significant $\rm [Fe/H]$
   variation ($\rm \Delta [Fe/H] = -0.05 \pm 0.04$, Fig. \ref{figPeriodFeHdifferentradii}) & All qualitatively consistent \\
   Sub-Neptunes & higher $\rm [Fe/H]$ at short period ($\rm \Delta [Fe/H] = 0.02 \pm 0.01$) & but quantitatively weaker \\
   Terrestrial planets & significantly higher $\rm [Fe/H]$ at short period ($\rm \Delta [Fe/H] = 0.07 \pm 0.01$) & \\ \hline
   \multicolumn{3}{c}{Sect. \ref{sec.eccen}: Orbital eccentricity vs. $\rm [Fe/H]$} \\ \hline
   Giant planets & positively-correlated (Fig. \ref{figGiantplaneteFeH}, \ref{figCDFFeHGPcategories}) & Qualitatively consistent but less significant \\
   Small planets with OGPs & positively-correlated (Fig. \ref{figSmallplanetssingle_eccFeH}, Eq. \ref{eqeccFeH_OGP}) & Quantitatively consistent \\
   Small planets w/o OGPs & not-correlated &  Inconsistent (Potential reasons in Sect. \ref{sec.eccen.SP}) \\ \hline

\end{tabular}}
\end{table*}   

\section{Summary}
\label{sec.Summary}

In this work, based on the synthetic planet population (Sect. \ref{sec:Synthesis}) generated by the Bern Generation III model (Sect. \ref{sec:model}), we perform a series of statistical analyses of planet properties as a function of stellar metallicity and compare with observational results. These results are summarised in Table \ref{tab:FeHcorrelations}.

We first investigate the dependence of the occurrence rates $F$ of planets with different sizes on the stellar metallicity (see Sect. \ref{sec.frequency}).
The main results are summarised as follows:
(1) The occurrence rates of the synthetic hot/warm/cold giant planets and Neptune-size planets show significantly positive correlations (Sect. \ref{sec.frequency.GP}- \ref{sec.frequency.Nep}, Fig. \ref{figFHJCJWJ_FeH_Synthesis}-\ref{figFHNCNWN_FeH_Synthesis}).
We perform a Bayesian analysis using an exponential model $F \propto 10^{\beta \times 10^{\rm [Fe/H]}}$.
The resulted $\beta$ are $\sim 1.3$ and $\sim 0.5-0.8$ for the giant planets and Neptune-size planets, respectively.
(2) For the Kepler planets with radii between $1-3.5 R_\oplus$, their occurrence rate $F$ as well as the fraction of stars hosting planets $\eta_{\rm system}$ and average multiplicity $k$ generally increase and then decrease with increasing $\rm [Fe/H]$ and the inflection point is $\sim 0.1$ dex (Sect. \ref{sec.frequency.Kep}, Fig. \ref{figFKep_FeH_Synthesis}). 
After excluding planetary systems with large planets ($R_{\rm p} > 3.5 R_\oplus$), the decreases in $F$ (as well as $k$ and $\eta_{\rm system}$) at $\rm [Fe/H]>0.1$ become weaker (see Fig. \ref{figFKep_FeH_SynthesisNonLP}), suggesting that the perturbation by large planets around metal-richer stars is an important mechanism causing the above inflection in the occurrence of Kepler-like planets.  
(3) For sub-Earths with $R_{\rm p} < 1 R_\oplus$, their occurrence rate $F$ as well as two decomposed factors $\eta_{\rm system}$ and $k$ all decline significantly with increasing $\rm [Fe/H]$ (Sect. \ref{sec.frequency.subEar}, Fig. \ref{figFSubEarth_FeH_Synthesis}). 
The occurrence rate-metallicity correlations derived from synthetic populations are generally consistent with the observational results for giant planets \citep[e.g.][]{2005ApJ...622.1102F,2010PASP..122..905J,2023PNAS..12004179C}, Neptune-size planets \citep[e.g.][]{2018PNAS..115..266D,2018AJ....155...89P,2022AJ....163..249C}, and Kepler planets \citep[e.g.][]{2015AJ....149...14W,2018AJ....155...89P,2019ApJ...873....8Z}.

We then study the radius valley morphology as a function of $\rm [Fe/H]$ using the synthetic population after applying the Kepler detection biases (Sect. \ref{sec.valley}).
We find that the radius valley deepens and the number ratio of super-Earths to sub-Neptunes decreases with increasing $\rm [Fe/H]$.
The fraction of Neptune-sized planets and the average radii
of planets above the radius valley are found to increase with [Fe/H], while the average radii of planets below the radius valley
($1-1.7 R_\oplus$) are broadly unchanged (Fig. \ref{figRadiusValleyFeH}- \ref{figRadiushistFeH}).
The evolutional trends of the five metrics are quantitatively consistent with those of \citet{2022AJ....163..249C} derived from the LAMOST-Gaia-Kepler sample.
The good matches between our nominal model and observations support that
the observed radius valley can be interpreted as the separation of two distinct planet populations, i.e. less massive, rocky super-Earths formed in situ and more massive water-rich sub-Neptunes formed ex-situ beyond the ice-line which then migrated inwards \citep{2020A&A...643L...1V,2024NatAs...8..463B}. 

In Sect. \ref{sec.distribution_radii}, we  explore how the period distributions of the synthetic planet populations depend on $\rm [Fe/H]$. 
We find that the orbital periods of synthetic planets are anti-correlated with $\rm [Fe/H]$ and planets with orbital periods less than 10 days are preferentially to be hosted by metal-richer stars with a difference in stellar metallicity $\Delta \rm  [Fe/H]$ of $ 0.05 \pm 0.01$ dex compared to those with longer periods (Fig. \ref{figPeriodFeHALL}).
The results also show that the anti-correlation (as well as $\Delta \rm [Fe/H]$) is most significant (largest as $0.07 \pm 0.01$ dex) for the terrestrial planets and becomes weaker with the increase of planet size (Fig. \ref{figPeriodFeHdifferentradii}-\ref{figDeltaFeHdifferentradii}).
The above synthetic results are qualitatively consistent with the observational results \citep{2014Natur.509..593B,2016AJ....152..187M}.

Finally, we investigate the correlations between the orbital eccentricity and host star $\rm [Fe/H]$.
We find that giant planets in eccentric orbits ($e>0.25)$ are preferentially to be hosted by metal-rich stars (Fig. \ref{figCDFFeHGPcategories}).
For small planets in single transiting systems, their eccentricities increase with stellar metallicities, while the eccentricities of small planets in multiple transiting systems are on average lower and have no significant correlation with $\rm [Fe/H]$ (Fig. \ref{figSmallplanetseFeH}).
Our results also show that metal-richer stars are more likely to be accompanied by (multiple) giant planets (top panels of Fig. \ref{figGantplaneteFeH_singlevsmultiple} and \ref{figSmallplanetssingle_eccFeH}) and thus the perturbations by giant planets are stronger to pump up the eccentricities of giant planets and small planets (bottom panels of Fig. \ref{figGantplaneteFeH_singlevsmultiple} and \ref{figSmallplanetssingle_eccFeH}).
Furthermore, the average masses of small planets increase significantly with $\rm [Fe/H]$ (Fig. \ref{figsmallplanetFeHmass}) and thus the self-interaction  should be stronger, further strengthening the eccentricity-metallicity correlation for small planets.

Overall, the synthetic population can generally reproduce the observed correlations between stellar metallicities and occurrence rates \citep[e.g.][]{2005ApJ...622.1102F,2010PASP..122..905J,2015AJ....149...14W,2018AJ....155...48W,2018AJ....155...89P,2019ApJ...873....8Z,2023PNAS..12004179C}, radius valley morphology \citep{2018MNRAS.478..197P,2022AJ....163..249C}, orbital periods \citep{2014Natur.509..593B,2016AJ....152..187M}, and eccentricity \citep{2013ApJ...767L..24D,2018ApJ...856...37B,2023AJ....165..125A}.
Nevertheless, the dependences of orbital period and eccentricity on the stellar metallicity derived from the synthetic sample are significantly weaker than those from observations.
Potential reasons could be: (1) the Generation III Bern model only considers the interactions between planets for the first 100 Myr and thus does not include the effect of long-term evolution of planets \citep{2021A&A...656A..69E}.
(2) The effects of stellar cluster environment at birth and binary companions are not included in the current model.

Future observations from both space and ground-based instruments
(e.g. GAIA, PLATO, NIRPS, ROMAN, ARIEL) will provide additional observational
datasets and relevant statistical analyses will provide more observational clues and constraints for planet formation and evolution.
Via planetary population synthesis, theoretical planet formation models can use this as an opportunity to perform not only qualitative but also quantitative comparisons in global ways between the synthetic population and observational datasets. The approach is particularly constraining if the same synthetic population is compared to datasets obtained from different observational methods like here where the NGPPS VII paper \citep{NGPPSVII} confronts the same population as used here with the constraints from the HARPS/Coralie GTO survey \citep{2011arXiv1109.2497M}. The reason is that different methods probe different model aspects.  
Such a quantitative multi-method approach will reveal the success
and limitations of our current understanding of planet formation more clearly and will eventually help to refine the theory of planet formation and evolution.

\section*{Acknowledgments}
This work is supported by the National Key R\&D Program of China (2024YFA1611803) and the National Natural Science Foundation of China (NSFC; grant No. 12273011, 12150009, 12403071). 
This research is also supported by the Excellence Cluster ORIGINS which is funded by the Deutsche Forschungsgemeinschaft (DFG, German Research Foundation) under Germany's Excelence Strategy – EXC-2094-390783311. 
C.M. and A.E. acknowledge the support from the Swiss National Science Foundation under grant 200021\_204847 ``PlanetsInTime''. R.B. acknowledges the financial support from DFG under Germany’s Excellence Strategy EXC 2181/1-390900948, Exploratory project EP 8.4 (the Heidelberg STRUCTURES Excellence Cluster).
Parts of this work has been carried out within the framework of the NCCR PlanetS supported by the Swiss National Science Foundation under grants 51NF40\_182901 and 51NF40\_205606.
J.-W.X. also acknowledges the support from the National Youth Talent Support Program.

\end{document}